\documentclass{aa}
\usepackage{graphicx}
\usepackage{enumerate}
\usepackage{txfonts}
\usepackage{url}
\usepackage{wasysym}
\usepackage{longtable}
\usepackage{array}
\usepackage{tikz}
\usepackage{pythonhighlight}
\usepackage[pdftitle={Pandora: A fast open-source exomoon transit detection algorithm}, colorlinks = true, breaklinks = true, citecolor = blue, linkcolor = blue, urlcolor = blue, pdfauthor = {Michael Hippke and Ren\'{e} Heller}]{hyperref}
\usepackage{esvect}

\DeclareMathAlphabet{\pazocal}{OMS}{zplm}{m}{n}

\begin{document} 

\title{Pandora: A fast open-source exomoon transit detection algorithm}
\titlerunning{Pandora: A fast open-source exomoon transit detection algorithm}

\author{
  Michael Hippke\inst{1,2}
  \and
  Ren{\'e} Heller\inst{3,4}
}

\institute{
   Sonneberg Observatory, Sternwartestra{\ss}e 32, 96515 Sonneberg, Germany\\
   \email{michael@hippke.org}
   \and
   Visiting Scholar, Breakthrough Listen Group, Berkeley SETI Research Center, Astronomy Department, UC Berkeley
   \and
   Max-Planck-Institut f\"ur Sonnensystemforschung, Justus-von-Liebig-Weg 3, 37077 G\"ottingen, Germany\\ \email{heller@mps.mpg.de}
   \and
   Institut f\"ur Astrophysik, Georg-August-Universit\"at G\"ottingen, Friedrich-Hund-Platz 1, 37077 G\"ottingen, Germany
}

   \date{Received 17 January 2022}

\abstract{We present {\tt Pandora}, a new software to model, detect, and characterize transits of extrasolar planets with moons in stellar photometric time series. {\tt Pandora} uses an analytical description of the transit light curve for both the planet and the moon in front of a star with atmospheric limb darkening and it covers all cases of mutual planet-moon eclipses during transit. The orbital motion of the star-planet-moon system is computed with a high accuracy as a nested Keplerian problem.
We have optimized {\tt Pandora} for computational speed to make it suitable for large-scale exomoon searches in the new era of space-based high-accuracy surveys. We demonstrate the usability of {\tt Pandora} for exomoon searches by first simulating a light curve with four transits of a hypothetical Jupiter with a giant Neptune-sized exomoon in a one-year orbit around a Sun-like star. The 10\,min cadence of the data matches that of the upcoming PLATO mission and the noise of 100 parts per million is dominated by photon noise, assuming a photometrically quiet, $m_V = 11$ Sun-like star for practicality. We recovered the simulated system parameters with the {\tt UltraNest} Bayesian inference package. The run-time of this search is about five hours on a standard computer. {\tt Pandora} is the first photodynamical open-source exomoon transit detection algorithm, implemented fully in the {\tt python} programming language and available for the community to join the search for exomoons.}

   \keywords{methods: data analysis -- occultations -- planets and satellites: detection -- stars: solar-type -- techniques: photometric
               }

   \maketitle

\section{Introduction}

While the search for planets beyond the Solar System (exoplanets) has greatly benefited from the transit method in search for periodic occultations of a star by its planets \citep[e.g.,][]{2013ApJS..204...24B,2019PNAS..116.9723Z}, no moon around any exoplanet has been securely discovered as of today. With over 3400 exoplanets found by the transit method, we naturally wonder if any exomoons could be detected around them. The detection of exomoon transits is very difficult due to the complex orbital motion, the occurrence of combined transits of an exoplanet with its moon (or moons), possible planet-moon eclipses, and potentially due to the small physical radii of exomoons, all of which make phase-folding approaches similar to those used for transiting exoplanets \citep{2002A&A...391..369K,2019A&A...623A..39H} inefficient. Indeed, methods like the orbital sampling effect \citep{2014ApJ...787...14H} or transit origami \citep{2021MNRAS.507.4120K} neglect about half of the tiny signal, details depending on the planet-moon orbital separation, and the transit geometry \citep{2021arXiv211104444H}.

Photodynamical modeling, on the other hand, is computationally extremely costly as it consists of up to 19 free parameters for inclined, eccentric orbits. \citet{2013ApJ...777..134K} report 49.7\,yr of CPU time for a single system using a similar computing infrastructure and multimodal nested sampling regression. \citet{2015ApJ...813...14K} report a mean value of 33,000\,hr of CPU time per planet-moon system on an unquantified number of AMD Opteron 6272 and 6282 SE processors. These times were recorded in a search for transiting exoplanets with exomoons in a sample of 41 light curves from quarters Q0-Q17 of the Kepler mission using the {\tt MultiNest} fitting software \citep{2009MNRAS.398.1601F}.

Beyond the {\tt LUNA} Fortran code \citep{2011MNRAS.416..689K} for transiting exoplanet-exomoon systems, other algorithms for the modeling of mutual transiting and eclipsing bodies exist. {\tt planetplanet} \citep{2017ApJ...851...94L}, software written in {\tt C} and wrapped in a {\tt python} interface, is taylored to model multitransiting exoplanet systems and it includes an analytical framework for the modeling of planet-planet mutual eclipses. Given the recent announcements of transiting exomoon candidates around Kepler-1625 \citep{2018AJ....155...36T} and Kepler-1708 \citep{Kipping2022} and the lack of available computer code to investigate these and future exomoon claims independently, an open-source exomoon transit detection algorithm is very desirable for the community.

Here we present \texttt{Pandora}, the first open-source exomoon transit detection algorithm\footnote{\url{www.github.com/hippke/pandora}} using a full photodynamical model. This state-of-the-art algorithm for exomoon transit modeling and searches requires just a few hours of CPU time for a Kepler-like light curve, which is about four orders of magnitude faster than previous algorithms, of which a factor of a few is due to today's faster CPUs. {\tt Pandora}'s speed gain is made possible through efficient computer code implementation in the {\tt python} programming language (e.g., using a just-in-time compilation of {\tt python} code using {\tt numba} \citep{numba}), a number of acceptable physical simplifications (e.g., eccentricity approximations), the symmetry of prograde and retrograde orbits, modern sampling algorithms, and other computational advances.

\section{Physical model}

\subsection{Parameterization}

At the heart of {\tt Pandora} sits a photodynamical model of a planet with a single moon that transits their common host star. Both the planet (with mass $M_{\rm p}$) and its moon (with mass $M_{\rm m}$) orbit their common center of mass (with mass $M_{\rm b}=M_{\rm p}+M_{\rm m}$) in Keplerian orbits. The orbit of the planet-moon barycenter around the star is parameterized by the midtime of the first barycenter transit in a given transit sequence ($t_0$), the orbital period of the barycenter around the star ($P_{\rm b}$), the semi-major axis of the barycenter orbit around the star ($a_{\rm b}$), and the transit impact parameter of the barycenter ($b_{\rm b}$). In general, the transit impact parameter can be parameterized in terms of the orbital inclination with respect to the line of sight as per $b_{\rm b} = a_{\rm b} \tan(\pi/2 - i_{\rm b}) / R_{\rm s}$, where $i_{\rm b}=\pi$ corresponds to an edge-on perspective.

Moving on to the parameterization of the planet-moon system, the planet has a radius $R_{\rm p}$ and the moon has a radius $R_{\rm m}$. In {\tt Pandora}, spatial dimensions are measured in units of the stellar radius and time is measured in units of days. As a consequence, the planet and moon radius, as well as the barycenter's orbital semi-major axis, are used as fractions of $R_{\rm s}$, that is, as ($R_{\rm p}/R_{\rm s}$), ($R_{\rm m}/R_{\rm s}$), and ($a_{\rm b}/R_{\rm s}$). By default, the orbit of $M_{\rm b}$ is assumed to have zero eccentricity ($e_{\rm b}=0$), but users can choose $e>0$ and then they would need to define the orientation of the periastron with respect to the line of sight ($\varpi_{\rm b}$) as well.

The orbits of the planet and the moon around their local center of mass are modeled as Keplerian orbits. By default, the planet-moon orbital eccentricity is zero, but users can parameterize nonzero eccentricities ($e_{\rm pm}$). In the default mode, the planet-moon orbit is modeled with an additional five parameters: orbital inclination $i_{\rm pm}$, longitude of the ascending node ($\Omega_{\rm pm}$), semimajor axis ($a_{\rm pm}$), orbital period ($P_{\rm pm}$), and time of periapsis passage ($\tau_{\rm pm}$). In {\tt Pandora}, $\tau_{\rm pm}$ is normalized with respect to $P_{\rm pm}$, that is $\tau_{\rm pm} \in (0,1)$. This choice is motivated by a significant boost in the convergence speed of our Monte Carlo sampling method (Sect.~\ref{sec:sampling}). For the calculation of the position of the planet and the moon on their elliptical orbit, however, we used $\tau P_{\rm pm}$. For eccentric orbits, the number of orbital elements increases to six, including $e_{\rm pm}$ and the argument of periapsis ($\omega_{\rm pm}$). Finally, the moon mass $M_{\rm m}$ is required to model the motion of the planet and moon around their joint barycenter.

The star is parameterized by $R_{\rm s}$ and two limb darkening coefficients ($u_1$, $u_2$) for the quadratic limb darkening law \citep{1977A&A....61..809M}. The quadratic limb darkening law is widely used in the exoplanet community because it reproduces stellar limb darkening sufficiently well for modern applications with space-based high-accuracy stellar photometry and because \citet{2002ApJ...580L.171M} derived an analytical solution to the resulting light curve for transiting planets. The stellar mass ($M_{\rm s}$) follows directly from the barycentric orbital mean motion $n_{\rm b}=2\pi/P_{\rm b}$ and $M_{\rm b}$ via Kepler's third law as $M_{\rm s}=n_{\rm b}^2 a_{\rm b}^3/G - M_{\rm b}$, where $G$ is the gravitational constant.


\subsection{Circumstellar orbital eccentricity}
\label{sec:circumstellar}

Our three-body system is a typical nested Keplerian system. For one thing, we assume that the orbits of the planet and the moon are not perturbed by the stellar gravitational force and that they orbit their common local barycenter in eccentric orbits. For another thing, we assume that the planet-moon barycenter follows a Keplerian orbit itself around the star.

\begin{figure}
\centering
\includegraphics[width=1.0\linewidth]{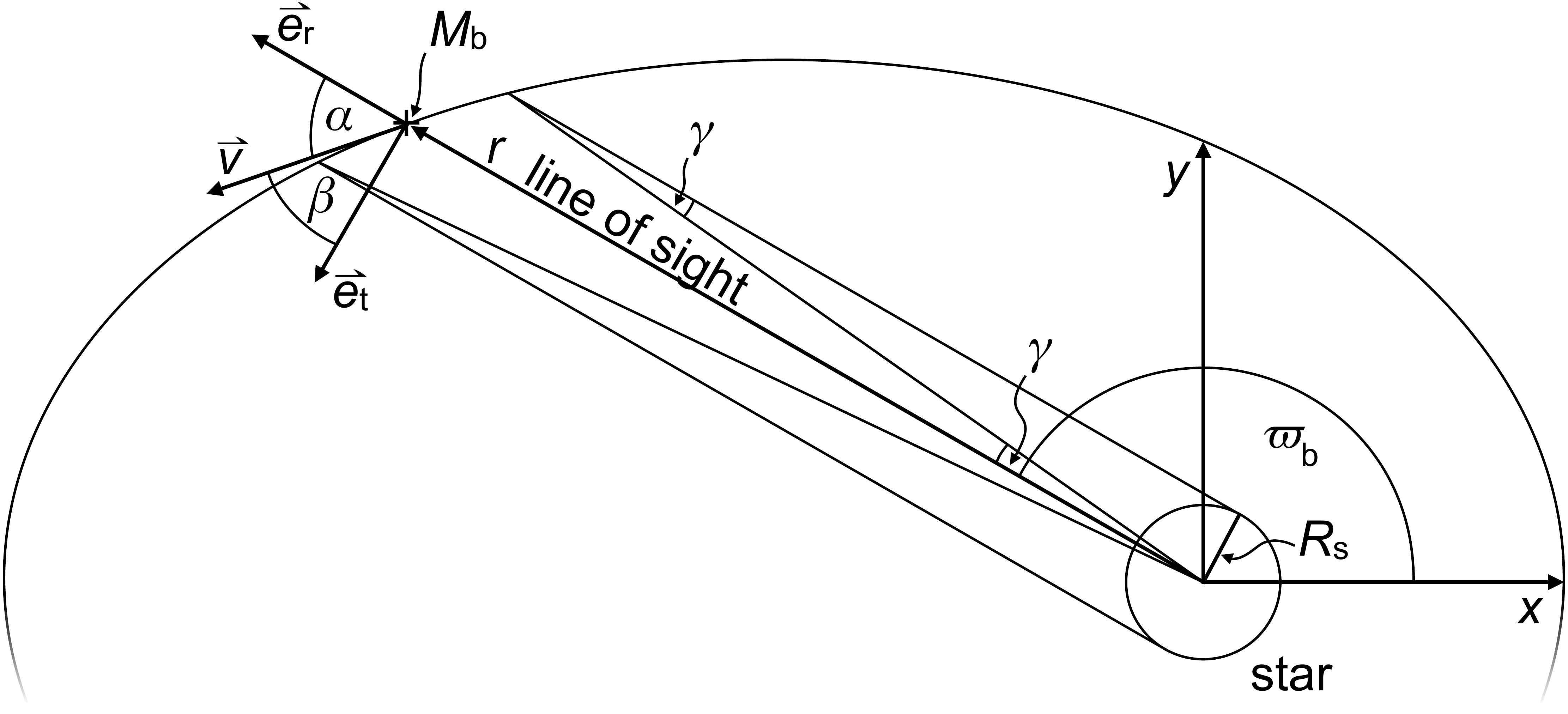}
\caption{Orbit geometry of the planet-moon barycenter ($M_{\rm b}$) on its eccentric circumstellar orbit as used in {\tt Pandora}. The situation depicted in this illustration corresponds to the midtransit time, when the true anomaly coincides with the orientation of the periastron ($\varpi_{\rm b}$). Nothing is to scale in this illustration. The bottom part of the ellipse has been spared for illustration purposes.}
\label{fig:ellipse}
\end{figure}

As for the orbit of $M_{\rm b}$, we are not interested in the full orbital revolution, but instead focus on an orbital section around the periodic transits. In our frame of reference (Fig.~\ref{fig:ellipse}), the line of the periapsis and the line of sight embraced an angle $\varpi_{\rm b}$, while the actual position of $M_{\rm b}$ on its orbit relative to the periapsis is given by the true anomaly ($f$). We plotted the orientation of the periapsis along the $x$-axis, in which case the orbital velocity of $M_{\rm b}$ is given as \citep{1999ssd..book.....M}

\begin{equation}
\vv{v} = (v_x, v_y) = \frac{n_{\rm b} \ a_{\rm b}} {\sqrt{1-e_{\rm b}^2}} {\Big (}-\sin(f), e+\cos(f) {\Big )}
.\end{equation}

\noindent
We are interested in the velocity component that is tangential with the celestial plane (perpendicular to the observer's line of sight), that is to say, the sky-projected velocity. We describe this direction by a unit vector $\vv{e}_{\rm t}$ and we introduced a unit vector $\vv{e}_{\rm r}$ for the (negative) radial velocity component:

\begin{align}
\vv{e}_{\rm r} =& \ {\Big (} \cos(f), \sin(f) {\Big )}\\
\vv{e}_{\rm t} =& \ {\Big (} -\sin(f), \cos(f) {\Big ).}
\end{align}

\noindent
The projection of $\vv{v}$ on $\vv{e}_{\rm r}$, which we refer to as $\vv{v}_{\rm r}$, is given as

\begin{equation}\label{eq:v_r}
\vv{v}_{\rm r} = | \vv{v} | \ \cos \sphericalangle (\vv{v},\vv{e}_{\rm r}) \ \vv{e}_{\rm r} = |\vv{v}| \cos(\alpha) \vv{e}_{\rm r} \ ,
\end{equation}

\noindent
where $\alpha \equiv \sphericalangle (\vv{v},\vv{e}_{\rm r})$ (Fig.~\ref{fig:ellipse}). The cosine of this angle between $\vv{v}$ and $\vv{e}_{\rm r}$ can be calculated as

\begin{equation}\label{eq:cos_al}
\cos(\alpha) = \frac{ \vv{v} \vv{e}_{\rm r} }{ |\vv{v}| \ |\vv{e}_{\rm r}| } = \frac{e_{\rm b} \sin(f)}{\sqrt{ \sin(f)^2 + (e_{\rm b} + \cos(f))^2 }} \ .
\end{equation}

\noindent
Plugging Eq.~\eqref{eq:cos_al} into Eq.~\eqref{eq:v_r} yields 

\begin{equation}\label{eq:v_r_tadaaa}
\vv{v}_{\rm r} = \frac{n_{\rm b} \ a_{\rm b} \ e_{\rm b}}{\sqrt{1-e_{\rm b}^2}} {\Big (} \sin(f)\cos(f), \sin(f)^2 {\Big )} \ .
\end{equation}

Analogously, projection of $\vv{v}$ on $\vv{e}_{\rm t}$, which we refer to as $\vv{v}_{\rm t}$, yields

\begin{equation}\label{eq:v_t}
\vv{v}_{\rm t} = | \vv{v} | \ \cos \sphericalangle (\vv{v},\vv{e}_{\rm t}) \ \vv{e}_{\rm t} = |\vv{v}| \cos(\beta) \vv{e}_{\rm t} \ ,
\end{equation}

\noindent
with

\begin{equation}\label{eq:cos_be}
\cos(\beta) = \frac{ \vv{v} \vv{e}_{\rm t} }{ |\vv{v}| \ |\vv{e}_{\rm t}| } = \frac{1 + e_{\rm b} \cos(f)}{\sqrt{ \sin(f)^2 + (e_{\rm b} + \cos(f))^2 }} \ .
\end{equation}

\noindent
Plugging Eq.~\eqref{eq:cos_be} into Eq.~\eqref{eq:v_t} yields 

\begin{equation}\label{eq:v_t_tadaa}
\vv{v}_{\rm t} = \frac{n_{\rm b} \ a_{\rm b}}{\sqrt{1-e_{\rm b}^2}} {\Big (}1 + e_{\rm b}\cos(f) {\Big )} {\Big (} -\sin(f), \cos(f) {\Big )} \ .
\end{equation}

\noindent
For our purpose of using the midtransit tangential velocity, we set $f = \varpi_{\rm b}$ and found

\begin{equation}\label{eq:v_t_abs}
|\vv{v}_{\rm t}| = \frac{n_{\rm b} \ a_{\rm b}}{\sqrt{1-e_{\rm b}^2}} {\Big (}1 + e_{\rm b}\cos(\varpi_{\rm b}) {\Big )}
,\end{equation}

\noindent
and using $p=2R_{\rm s}\sqrt{1-b_{\rm b}^2}$ as the transit path across the star, we constructed the transit duration of a point-like $M_{\rm b}$ as

\begin{equation}\label{eq:t_dur}
d = \frac{p}{|\vv{v}_{\rm t}|} = \frac{2R_{\rm s}\sqrt{1-b_{\rm b}^2} \sqrt{1-e_{\rm b}^2} }{n_{\rm b} \ a_{\rm b} {\Big (}1 + e_{\rm b}\cos(\varpi_{\rm b}) {\Big )} }
.\end{equation}

\noindent
In Appendix~\ref{sec:eccentricity_error} we extend these considerations to the tangential transit velocity during ingress and egress and demonstrate that our assumption of constant in-transit velocity of the planet-moon barycenter produces errors of much less than 1\,part per million (ppm) even under very unfavorable conditions. Most importantly, Eq.~\eqref{eq:t_dur} is computationally much faster than using a Kepler solver for the barycentric orbital motion around the star.

\subsection{Planet and moon orbits around the local barycenter}
\label{sec:planetmoon}

The relative positions of the planet and its moon with respect to their local center of mass are calculated using a 2D Kepler solver. By default, {\tt Pandora} assumes a circular orbit as the standard parameterization for large-scale exomoon surveys in hundreds to thousands of light curves with known transiting exoplanets. With $e_{\rm pm}=0$ and $\varpi_{\rm pm}=0$, the 2D Kepler solver requires four of the usual six Keplerian elements plus the time of any given data point as input parameters. It then returns the positions of both bodies using the analytical solution of the circular Kepler orbit and without any costly iterative approximations (Appendix~\ref{sec:trigo}). Hence, in {\tt Pandora}'s default mode, the CPU run time of the 2D Kepler solver is as costly as any analytic approximation.

For eccentric planet-moon orbits, which might be interesting to study in more detail for potential exomoon candidates or to examine eclipse phenomena, etc., {\tt Pandora} solves the Kepler equation using the approach by \citet{1995CeMDA..63..101M} similarly to the implementation in {\tt PyAstronomy} \citep{pya}. The solution is built on a fifth-order refinement of a cubic equation. It can be executed in a single iteration and requires the calculation of four transcendental functions.

The star is at the origin of our ($x,y$) coordinate system and all distances are measured in stellar radii. The distance of the center of any transiting body (planet or moon) from the center of the stellar disk, which is required for the flux calculations, is thus simply given as $\sqrt{x^2+y^2}$.

Before calculating any data points for the Kepler ellipse, however, {\tt Pandora} estimates if the planet-moon barycenter is sufficiently close to the stellar disk for any transits to occur in the first place. If the planet-moon barycenter is farther away from the origin of the coordinate system than the sum of the stellar radius and the maximum possible deflection of the moon from the planet, $(1+a_{\rm pm})e_{\rm pm}$, then occultations are geometrically impossible. In this case, no calculations are made and the flux in the model light curve is set to 1. This saves CPU time out of transit.

{\tt Pandora}'s planet-moon orbit module has a maximum throughput between 14\,million (if $e>0$) and 21\,million (if $e=0$) data points per second per core on an Intel i7-1185G7 processor.

\begin{table}
\caption{Model parameters}
\label{table:params}
\centering
\begin{tabular}{lll}
\hline\hline
Star          & Description                      & Unit        \\ 
\hline
$R_{\rm s}$   & Stellar radius                   & m           \\
$u_1$, $u_2$  & Limb darkening                   & 1      \\
\hline
\hline
Barycenter \\
\hline
$P_{\rm b}$         & Period                           & days        \\
$t_{\rm 0,b}$       & Time of inferior conjunction     & days        \\
$a_{\rm b}$         & Semimajor axis                   & $R_{\odot}$ \\
$b_{\rm b}$         & Transit impact parameter                 & 1      \\
$e_{\rm b}$         & Eccentricity                     & 1      \\
$\varpi_{\rm b}$    & Argument of periapsis            & deg         \\
\hline
\hline
Planet \\
\hline
$R_{\rm p}$         & Radius                           & $R_{\odot}$ \\
$M_{\rm p}$         & Mass                             & kg          \\
\hline
\hline
Moon \\
\hline
$R_{\rm m}$         & Radius                           & $R_{\odot}$ \\
$P_{\rm pm}$            & Period                           & days        \\
$i_{\rm pm}$            & Inclination                      & deg         \\
$\Omega_{\rm pm}$       & Longitude of the ascending node  & deg         \\
$\varpi_{\rm pm}$       & Argument of periapsis            & deg         \\
$e_{\rm pm}$            & Eccentricity                     & deg         \\
$\tau_{\rm pm}$         & Time of periapsis passage        & 1      \\
$M_{\rm m}$   & Mass           & kg      \\
\hline
\end{tabular}
\end{table}

\subsection{Stellar flux and limb darkening}
\label{sec:flux}

To calculate the apparent stellar flux during transits of the planet-moon pair, we used the analytical equations of \citet{2002ApJ...580L.171M}. Our code was adapted from \texttt{PyTransit} \citep{2015MNRAS.450.3233P}, which was released under the GPL open source license. {\tt Pandora}'s transit module requires the distance between the centers of the star and the transiting body (provided as a series of values by {\tt Pandora}'s Kepler ellipse module), the radius of the transiting circle in units of the stellar radius, and the quadratic limb darkening parameters. It returns the series of flux values where an unocculted star is set to unity.

To speed up calculations, {\tt Pandora} uses a hybrid occultation model, where the small-body approximation \citep{2002ApJ...580L.171M} where constant stellar limb darkening behind the occulted area is assumed whenever appropriate. The error of the small-body approximation is most pronounced during transit ingress and egress. But even then, it is typically on the order of 1 ppm, as long as $R_{\rm p}/R_{\rm s}<0.01$, and it only accounts for about 1\,\% of the transit duration, with details depending on the transit impact parameter. For bodies with sizes $0.01 < R_{\rm p}/R_{\rm s} < 0.05$, {\tt Pandora} employs a hybrid model. It calculates two exact values and linearly interpolates intermediate values from the small-planet approximation. The resulting model fluxes are accurate to $<1\,$ppm in all cases (see Appendix \ref{appendix:small_body_approxi}).

In case of fixed limb-darkening parameters during sampling, {\tt Pandora} creates a lookup table of occultation values before the sampling commences. This procedure calculates a 2D grid of occultation values covering planet-to-star radius ratios $0 < (R_{\rm p}/R_{\rm s}) < 0.2$ and distances from the stellar center $0 < z < 1 + (R_{\rm p}/R_{\rm s})$ with $300$ values along each axis, which we verified to result in errors $<1\,$ppm. Afterwards, the occultation values are calculated for each model from this rectilinear 2D grid with a bilinear interpolation from their nearest neighbors. This procedure only requires four floats to be obtained from memory, and at most eight multiplications, one division, and thirteen additions per data point.

The throughput of {\tt Pandora}'s transit algorithm is 16\,million data points per second per core on an Intel i7-1185G7 processor for $R_{\rm p}/R_{\rm s} > 0.05$, 48\,million for $R_{\rm p}/R_{\rm s} = 0.02$, and 110\,million for $R_{\rm p}/R_{\rm s} < 0.01$. For cached values (in the case of fixed limb darkening), the maximum throughput is 200 million points per second, which is more than ten times the number of a regular calculation as per \citet{2002ApJ...580L.171M}. In combination with our approximation for the transit duration in eccentric orbits (Sect.~\ref{sec:circumstellar}), fitting planet-only models with {\tt Pandora} is up to fifty times faster than {\tt batman} \citep{2015PASP..127.1161K}. We show a performance comparison of all methods as a function of the number of data points in Appendix~\ref{fig:occult_speed}.

\subsection{Planet-moon eclipses}
\label{sec:eclipses}

As long as the planet and the moon transit the star without mutual eclipses, {\tt Pandora} treats these synchronous transits separately, as if the star were occulted by two independent spherical bodies. If, however, the circles of the planet and the moon overlap during ingress or egress of at least one of the two bodies, then things become more complicated. Hence, planet-moon eclipses during ingress or egress require special treatment.

In his derivation of the equation for the area of common overlap of three circles, \citet{Fewell2006} identified nine cases for the intersection of three circles in a plane. His case (1) presents a circular triangle, for which an analytic solution did not exist until then. \citet{2011MNRAS.416..689K} extended this system to 27 cases for various star-planet-moon arrangements plus specific subcases, all of which are distinguished by specific conditions of the relative distances and radii of the three circles. Our approach with {\tt Pandora} is much more pragmatic, though still computationally affordable and numerically accurate to arbitrary levels. {\tt Pandora} distinguishes only between two cases of a planet-moon eclipse and still covers all possible cases of previous studies.

\subsubsection{Planet-moon eclipses and no contact with the stellar limb}

{\tt Pandora}'s eclipse case (1) is called if the planet and the moon are both fully on the stellar disk during a mutual event, that is, if

\begin{align}
1 - \sqrt{x_{\rm m}^2 + y_{\rm m}^2} < R_{\rm m} \hspace{0.4cm} {\rm and} \hspace{0.4cm} 1 - \sqrt{x_{\rm p}^2 + y_{\rm p}^2} < R_{\rm p} \hspace{0.4cm} {\rm ,}
\end{align}

\noindent
while the distance between their centers ($d_{\rm pm}$) is smaller than the sum of their radii,

\begin{align}
d_{\rm pm}<R_{\rm P}+R_{\rm m} \hspace{0.4cm} {\rm .}
\end{align}

\noindent
{\tt Pandora} ignores the R\o{}mer effect for exomoons \citep[dubbed ``transit timing dichotomy'';][]{2014ApJ...796L...1H}, which is on the order of a few seconds for realistic planet-moon systems. As a consequence, it is irrelevant for the computation light curve whether the moon is physically located in front of the planet or behind the planet during eclipse.

The decrease in the stellar apparent brightness due to the planet and the moon were initially calculated separately. Then we analytically calculated the area of the asymmetric lens defined by the intersection of the two circles \citep[see Eq.~13 in][]{2016ApJ...820...88H} and, assuming that the stellar limb darkening is constant under this small area, compensated for the twofold consideration of this small area  analytically in the initial calculation. The resulting error is $<0.1$\,ppm. The throughput of {\tt Pandora}'s eclipse case (1) is 2\,billion data points per second per core.

\begin{figure*}
\centering
\includegraphics[width=.49\linewidth]{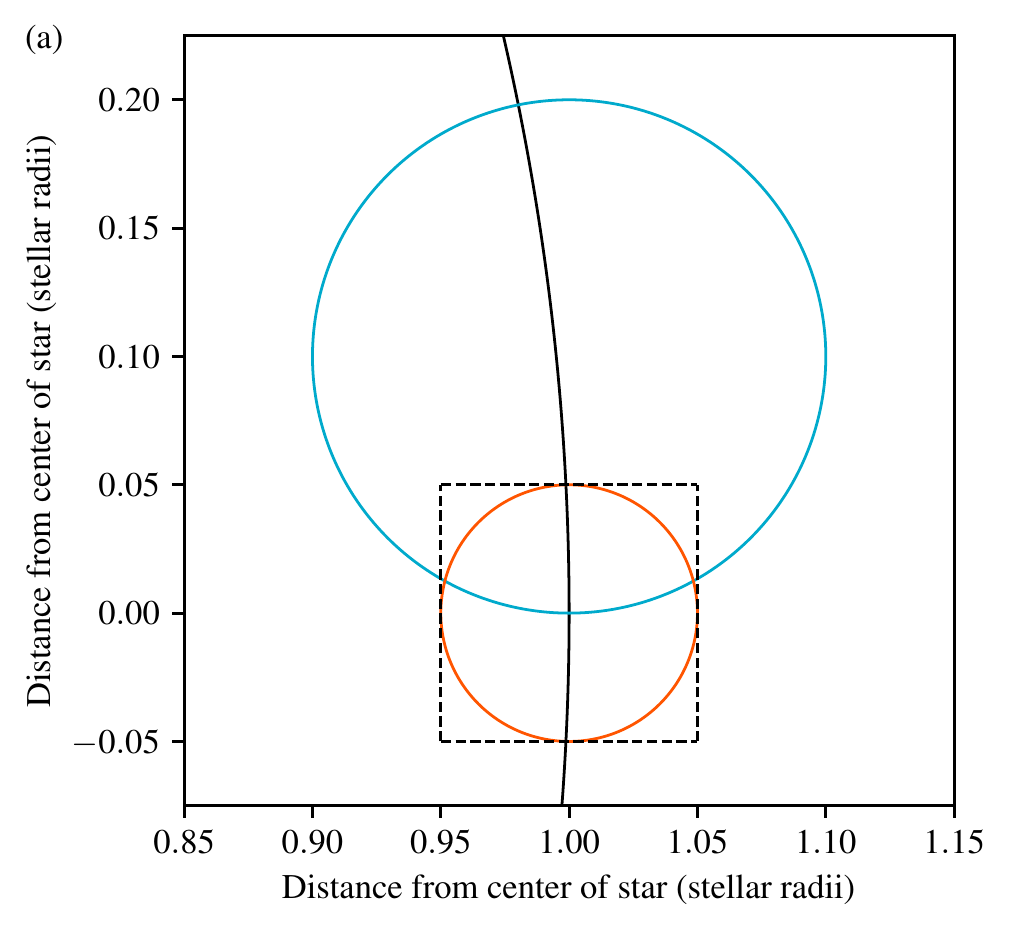}
\includegraphics[width=.45\linewidth]{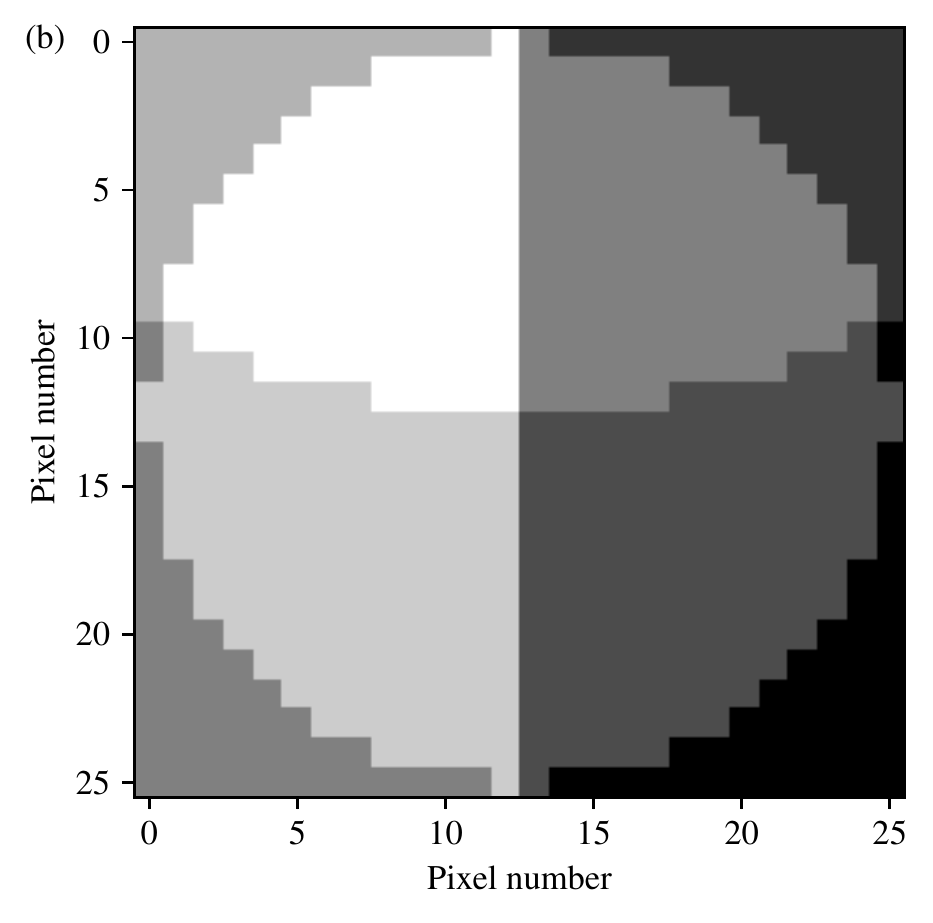}
\caption{Planet-moon eclipse in contact with the stellar limb. (a) The planet (blue circle) and the moon (orange circle) are both in front of the star (black circle segment) and in eclipse. (b) Numerical pixel grid representing the moon. Different areas of overlap are shaded in arbitrary colors. In this specific configuration, the white area in the upper left region of the image is determined by pixel area summation and then compensated for in the light curve. The right half of the moon is not transiting.}
\label{fig:pixelart}
\end{figure*}

\subsubsection{Planet-moon eclipses in contact to the stellar limb}
\label{sec:eclipsecase2}

{\tt Pandora}'s eclipse case (2) occurs if at least one of the transiting bodies touches the stellar limb during a planet-moon eclipse. Then the area of common overlap of the three circles is integrated numerically and the correction of the light curve is performed as for {\tt Pandora}'s eclipse case (1) assuming constant stellar limb darkening for this small area.

This numerical treatment is illustrated in Fig.~\ref{fig:pixelart}. Panel (a) shows a sketch of three overlapping circles, where the black line depicts the stellar limb, the blue circle the planet, and the orange circle the moon. The dashed rectangle denotes the region shown in the zoom in panel (a), where we show a pixelated representation of the moon. In this particular case, the moon has a diameter of 25 pixels. In Appendix~\ref{sec:eclipse_error} we demonstrate that this resolution is sufficient to reduce the resulting numerical error below 1\,ppm. The different colors for the different areas of overlap in Fig.~\ref{fig:pixelart}(b) were chosen solely for illustrative purposes. In particular, the white area in the upper left region is the area of common overlap that {\tt Pandora}'s numerical eclipse module determines and compensates for in this specific example.

The throughput of {\tt Pandora}'s eclipse case (2) is 0.25\,million data points per second and per core. This is relatively slow compared to {\tt Pandora}'s out-of-eclipse throughput of between 16 and 200 million data points per second and per core (Sect.~\ref{sec:flux}). However, since planet-moon eclipses are only relevant for planet-moon systems with orbital inclinations less than a few degrees to the line of sight and since they are usually very short and only relevant to a few data points, this {\tt Pandora} module only accounts for a small fraction of {\tt Pandora}'s total computing time.

\subsection{Supersampling}

If the exposure time of a light curve is longer than the time scale of the variation due to the signal, then this signal becomes temporally smeared \citep{2010MNRAS.408.1758K}. In {\tt Pandora}, this temporal smearing can be compensated for by using supersampling of the light curve, which naturally comes at a higher computational cost and which is thus optional for users.

The supersampling factor can be set to any integer $\geq 1$, where a factor of one corresponds to no supersampling, and a factor of five to seven is a sensible upper limit beyond which additional accuracy gains become marginal. The supersampling factor has a near-linear influence on the model calculation time.

\section{Results}
\label{sec:sampling}

\subsection{{\tt Pandora} software}

{\tt Pandora} is a pure {\tt python} software implementation of the modules describes in the previous sections. As for its working procedure, it first calculates the position of the planet-moon barycenter for each timestamp $t$ (Sect.~\ref{sec:circumstellar}). Then the Kepler ellipse for the planet-moon binary is computed in order to determine the $(x, y)$ positions of the two bodies with respect to the stellar disk (Sect.~\ref{sec:planetmoon}). With these positions, the occultation function is called to construct the flux drop resulting by both bodies separately (Sect.~\ref{sec:flux}). Finally, a check for mutual eclipses and a correction to the flux is made if necessary (Sect.~\ref{sec:eclipses}).

\begin{figure*}
\centering
\includegraphics[width=1.0\linewidth]{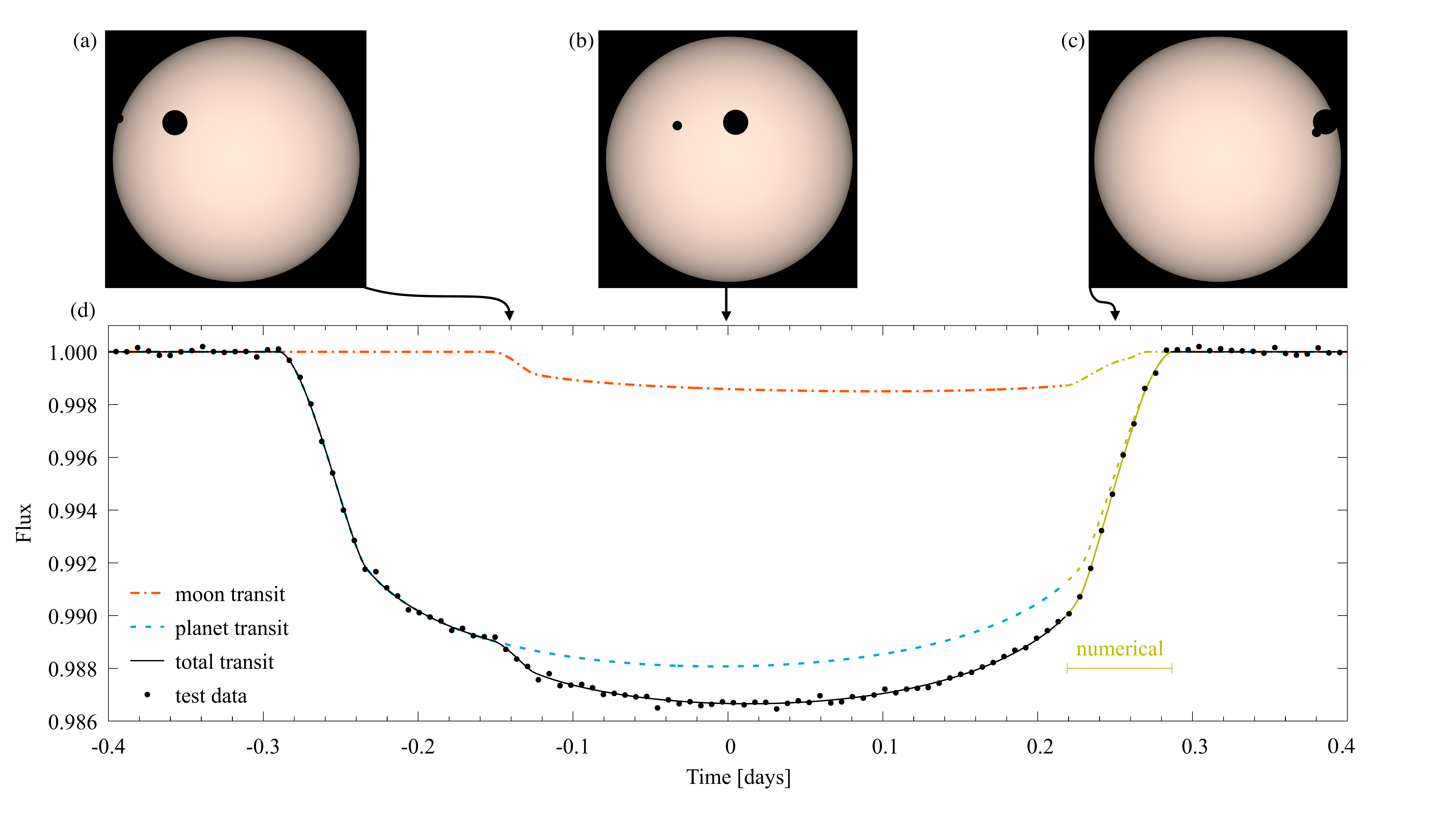}
\caption{Output demo of {\tt Pandora} for a system of a Sun-like star, a Jupiter-sized planet in a one-year orbit around that star, and a Neptune-sized moon in a 1.28\,d orbit around the giant planet (for details see Sect.~\ref{sec:output}). (a)-(c) Video renderings of the transit. (d) Light curve. The dotted-dashed orange line shows the transit light curve of the exomoon and the dashed blue line shows the transit light curve of the exoplanet using the analytical solution in both cases. The yellow parts of the moon and planet light curves illustrate numerical simulations. The solid black line shows the combined model. Black dots show a simulated observation roughly representative of an $m_V\,\sim\,11$ Sun-like star from the PLATO mission. Digital star colors are from \citet{2021AN....342..578H}.}
\label{fig:pandora}
\end{figure*}

In total, {\tt Pandora} consists of 640 source lines of code (SLOCs) in the {\tt python} programming language. This includes the main module and all functions. In addition, {\tt Pandora} uses a class-based wrapper function (56 SLOCs) and a video generator (83 SLOCs). {\tt Pandora} only depends on the {\tt numpy} library \citep{harris2020array} and the {\tt numba} just-in-time compiler. If {\tt Pandora}'s own occultation code was replaced, for example, with a call to {\tt PyTransit} to obtain the occultation values, only 318 SLOCs would remain. This makes {\tt Pandora} a comparably small code base. Most other packages have a lot more lines of code, such as {\tt W{\={o}}tan} \citep[1406 SLOCs]{2019AJ....158..143H}, {\tt TLS} \citep[1689 SLOCs]{2019A&A...623A..39H}, or {\tt batman} \citep[256 SLOCs in {\tt python} and 976 SLOCs in {\tt C}]{2015PASP..127.1161K}, not counting test files. A smaller code base has the advantage of making a package easier to understand, debug, and extend. While being concise, {\tt Pandora} is also readable with clear variable names, a logical modular structure, and comments which explain each code line if necessary.

For debugging purposes and to test the correct implementation of our model in {\tt Pandora}, we compared single-body occultations with limb darkening and super-sampling against {\tt PyTransit} and {\tt batman}, and the planet-moon orbit against {\tt PyAstronomy}, including required coordinate transformation to solve for the Kepler ellipse. All test cases agreed to within 0.01\,ppm per data point. We also created a series of test light curves and the corresponding video animations, and stepped through these frame by frame to validate each data point.

\texttt{Pandora} is concise and modular. Its individual components can be readily understood, tested, and revised as needed. \texttt{Pandora} is an open source code and we invite the community to run their own experiments and submit feature requests, ideas for improvement, and bug reports directly to the GitHub tracker\footnote{\href{https://github.com/hippke/Pandora/issues}{https://github.com/hippke/Pandora/issues}}. We have previously made very positive experience with community contributions to {\tt TLS}  and {\tt W{\={o}}tan}, for which a total of about 30 improved versions have been released over the course of the first two years based on community pull requests (i.e., code contributions), bug reports (usually for small issues and edge cases), and general modifications to improve the workflow.

{\tt Pandora} can generate light curves from custom orbital geometries. These light curves can have arbitrary time sampling and the data quality can be artificially reduced with white noise, for which the amplitude can be set by the users. {\tt Pandora} can also create transit animation videos including a star with realistic limb darkening and color \citep[as per][]{2021AN....342..578H} at a custom resolution and frame rate. The light curves can be used for exomoon searches. The videos can be useful for education and outreach. {\tt Pandora} can also output the spatial coordinates, which we found useful for testing and debugging. They could also be used to compare to observational radial velocity data.

\subsection{{\tt Pandora} output}
\label{sec:output}

In Fig.~\ref{fig:pandora} we show an example of an exoplanet-exomoon system modeled with {\tt Pandora}. The star is assumed to have a solar radius and solar-type limb darkening, with $u_1=0.4089$ and $u_2=0.2556$ as per \citet{2011A&A...529A..75C} for a star with a surface gravity of $\log(g)=4.5$, and an effective temperature of $T_{\rm eff}=5750$\,K, a metallicity of [M/H]~=~0.0, and zero microturbulent velocity. The planet-moon barycenter has an orbital period of 365.25\,d around the star and a transit impact parameter $b=0.3$. The planet is as large and as heavy as Jupiter. The moon is as large and as heavy as Neptune, thus ($M_{\rm m}/M_{\rm p}=0.05395$), with an orbital period of 1.28\,d around the planet-moon barycenter.

This period, which implies an orbital semi-major axis of $5\,R_{\rm p}$, ensures that the moon is beyond the Roche radius of the planet, which we determined to be at about $2.3\,R_{\rm p}$ \citep{2010A&A...521A..76W} under the assumption of fluid-like objects. For comparison, Io, the innermost of the Galilean moons around those of Jupiter, is at roughly $6.1\,R_{\rm p}$ around Jupiter. The planet-moon orbit is slightly inclined with respect to the line of sight ($i_{\rm pm}=80^\circ$). We do not claim that our model system is particularly representative of real exomoons. However, it is physically possible and well-suited to illustrate the features of {\tt Pandora}.

In addition to the model light curve, we generated a simulated observation that is roughly representative of an idealized, photometrically quite, Sun-like star as if it were observed by the PLATO mission. PLATO is expected to launch in 2026 in search for transiting exoplanets around bright stars \citep{2014ExA....38..249R}. We chose a cadence of 10\,min as will be used for PLATO's P5 stellar sample of ${\geq}\,245,000$ dwarf and subdwarf stars of spectral types F5-K7 with apparent visual magnitudes $11\,\leq\,m_V\,\leq\,13$. We added a white noise component to each data point that is randomly drawn from a normal distribution with a standard deviation of 100\,ppm. This noise component is roughly representative of a photometrically quiet Sun-like star from PLATO's P5 sample\footnote{PLATO Definition Study Report, \href{https://sci.esa.int/s/8rPyPew}{https://sci.esa.int/s/8rPyPew}}, but it ignores any kind of astrophysical variability. Specifically, the photon noise from this test star would correspond to an apparent visual magnitude $m_V \sim 11$, though the details of the photon noise floor also depend on the actual number of cameras (6, 12, 18, or 24) with which the target is observed. These simulations assume ideal conditions for the purpose of illustrating {\tt Pandora's} functionality. For details on the actual detectability of exoplanet transits with PLATO in the presence of stellar noise, see \citet{HHS2022}.

Figure~\ref{fig:pandora}(a) shows a snapshot of the video animation generated with {\tt Pandora} that is taken during midingress of the moon. We configured the system in such a way that the planet is already in transit when the moon begins its transit. Figure~\ref{fig:pandora}(b) refers to the midtransit time of the planet-moon barycenter, at which point the planet and the moon exhibit near-maximum tangential deflection with respect to the line of sight. In Fig.~\ref{fig:pandora}(c) the planet is in egress during an eclipse, in which case {\tt Pandora} switches from the analytical treatment of the light curve to numerical simulations. Figure~\ref{fig:pandora}(d) presents the resulting transit light curve model (solid black line) and simulated PLATO observation (black dots) as well as a breakdown of the contributions from the moon (dotted-dashed orange line) and the planet (dashed blue line). Table~\ref{tab:runtime} summarizes the computing time that {\tt Pandora} spent for the various processes.

\begin{figure*}
\centering
\includegraphics[width=1.0\linewidth]{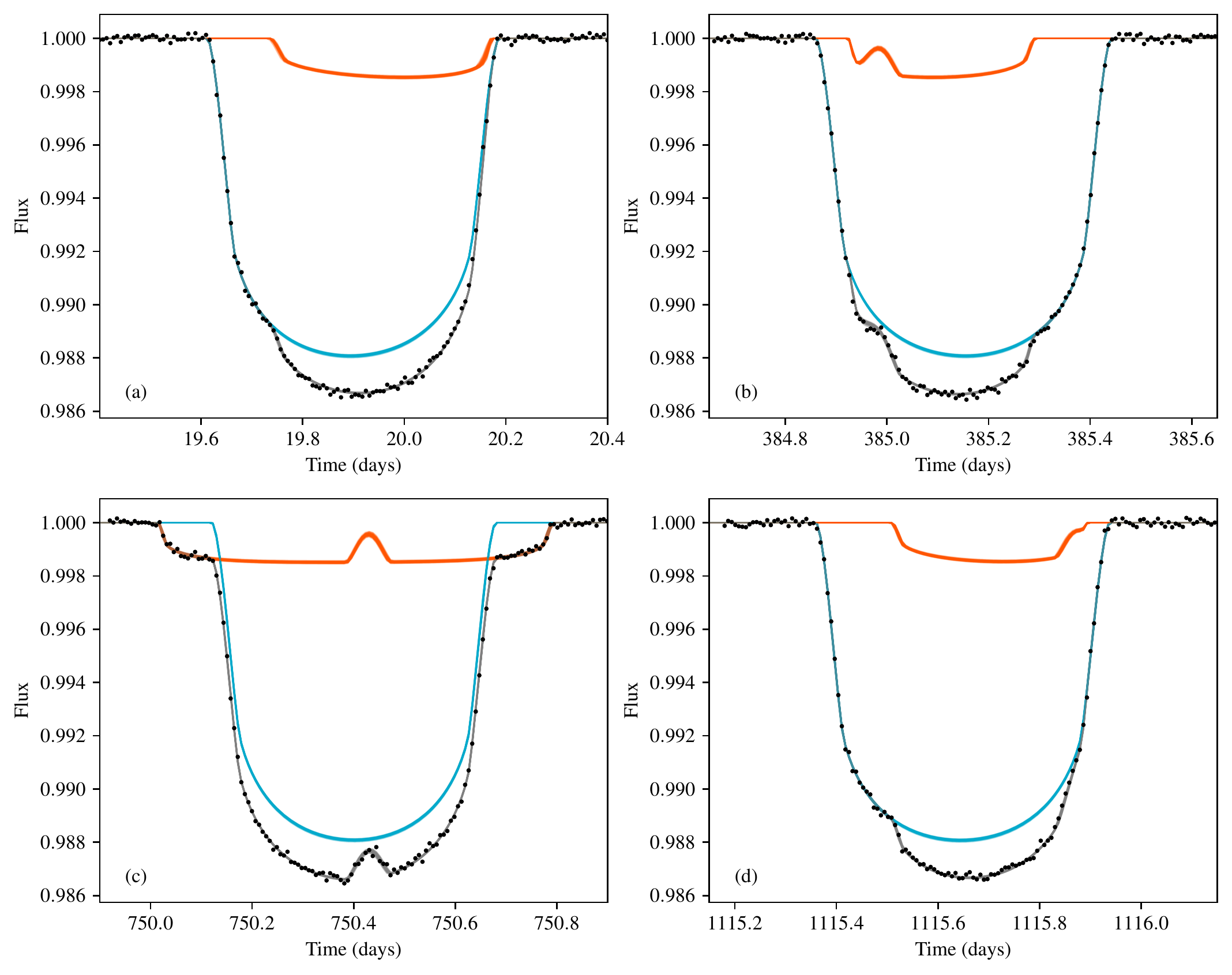}
\caption{Simulated planet-moon observations (black points) with a noise amplitude of 100\,ppm per data point and a 10\,min cadence, roughly corresponding to an idealized observation of an $m_V\,\sim\,11$ photometrically quiet Sun-like star in PLATO's P5 sample. The system is the same as in Fig.~\ref{fig:pandora} (see Sect.~\ref{sec:output} for details). (a) Same transit as in Fig.~\ref{fig:pandora}(d), but with a new realization of the random noise. From our {\tt UltraNest} recovery, we selected 100 model parameters from the posteriors and we show their light curves for the moon (orange), planet (blue), and total (gray) light curves.}
\label{fig:LC_retrieval}
\end{figure*}

\begin{figure*}
\centering
\includegraphics[width=\linewidth]{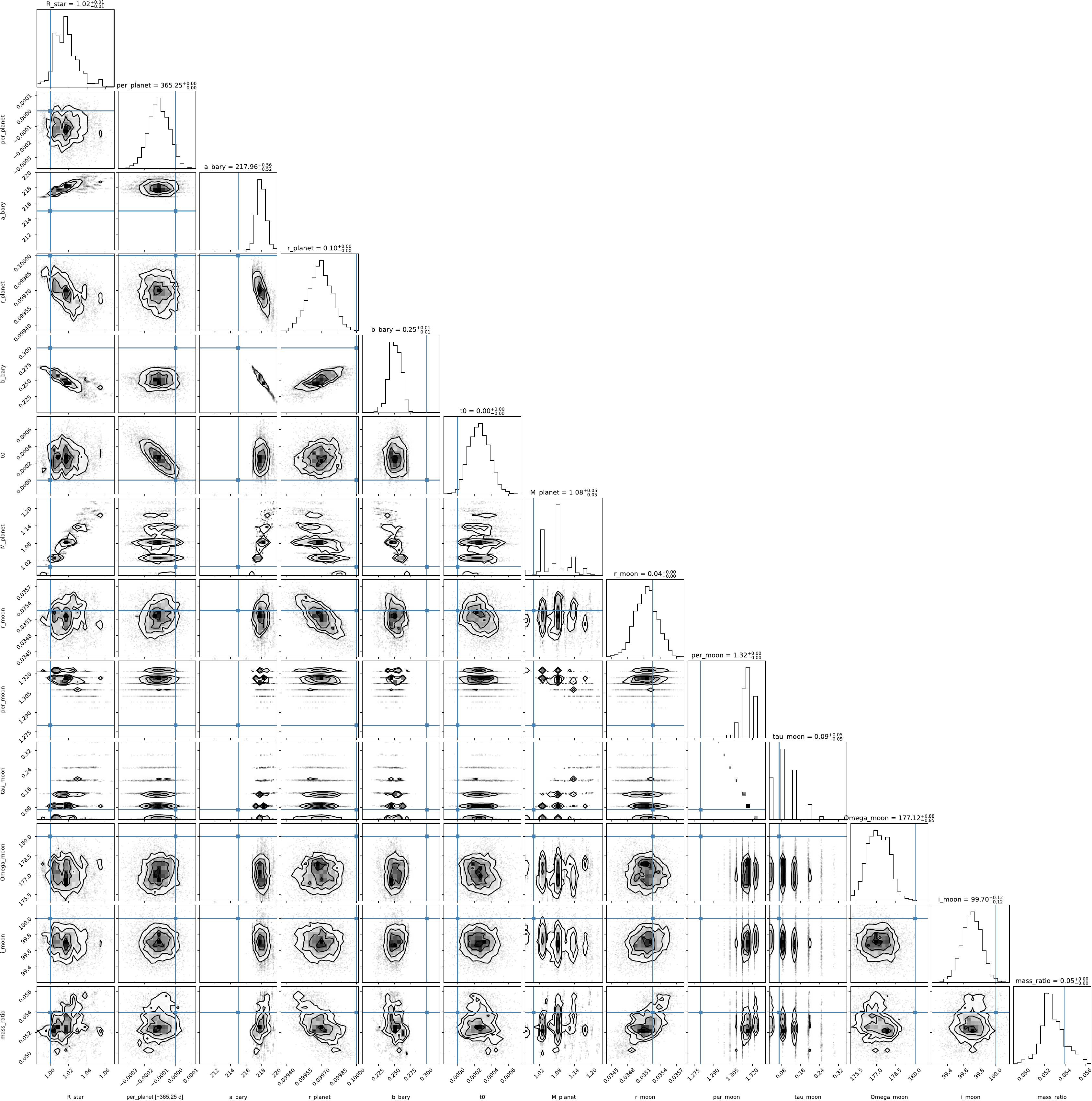}
\caption{Corner plot from the {\tt UltraNest} retrieval of the system shown in Fig.~\ref{fig:pandora}, but with four transit epochs. We note that $R_{\rm s}$ is given in units of $R_\odot$, $a_{\rm b}$ in units of AU, and $M_{\rm p}$ in units of Jupiter masses. Due to the small number of transits, the posterior is multimodal for combinations of $P_{\rm pm}$ and $\tau_{\rm pm}$. Additional transits would constrain the system much better.}
\label{fig:cornerplot}
\end{figure*}

\begin{table}[h]
    \caption{Run time breakdown of {\tt Pandora} for the example system described in Sect.~\ref{sec:output}.}
    \centering
    \begin{tabular}{l|c}
        \hline\hline
        Module & Process  \\
        \hline
        Moon transit               & 7\,\%   \\
        Overhead                   & 9\,\%   \\
        Coordinate transformations & 15\,\%  \\
        Eclipses                   & 22\,\%  \\
        Kepler ellipse             & 23\,\%  \\
        Planet transit             & 24\,\%  \\
        \hline
    \end{tabular}\\
    \tablefoot{For the analytical part of the light curve, the moon transit was computed with the small body approximation and quadratic limb darkening \citep{2002ApJ...580L.171M}. For the numerical transit simulations, which {\tt Pandora} uses if planet-moon eclipses occur during ingress or egress of one of the two bodies, the moon was modeled with a diameter of 25 pixels.}
    \label{tab:runtime}
\end{table}

\subsection{Simulated exomoon search}

To demonstrate {\tt Pandora}'s performance as an exomoon search tool, we first simulated a single light curve based on the same star-planet-moon system as described in Sect.~\ref{sec:output}, but we extended the data set to four transit epochs. The resulting mock observations with PLATO are shown with black dots in Fig.~\ref{fig:LC_retrieval}.

Then we fit the simulated PLATO data using \texttt{UltraNest} \citep{Buchner2021}, a general-purpose Bayesian inference package for parameter estimation and model comparison. It allows one to fit arbitrary models for multimodal or non-Gaussian parameter spaces as can be expected for exoplanet-exomoon systems due to alias effects in the orbital period of the planet-moon system. \texttt{UltraNest} also features computer parallelization and the resumption of incomplete runs.

We chose uninformed flat priors for all parameters, but fixed limb darkening to the injected values. All eccentricities were set to zero. As for the sampling method, we ran the {\tt StepSampler} of {\tt UltraNest} with 4000 accepted steps until the sample was considered independent, and we let the sampler decide the number of currently active walkers, which typically resulted in hundreds to thousands of active walkers.

Upon convergence after about 5\,hr of run time on a single core of an Intel i7-1185G7 processor, we found the solution shown in the corner plot of Fig.~\ref{fig:cornerplot}. The sampler performed roughly 250\,million model evaluations in total, which equates to an average of about $13,900$ models and log-likelihood calculations per second. We expect linear scaling with more cores and more light curves due to the independence of a search in available data, for example when searching for moons in Kepler or TESS data.

In Fig.~\ref{fig:LC_retrieval} we show 100 randomly drawn models from the posterior distribution. The transparent black lines illustrate the total model, the transparent orange lines refer to the transit contribution by the exomoon, and the transparent blue lines show the stellar dimming caused by the exoplanet.

As for the recovery of the injected planet-moon transit, the model parameters are indicated with blue dots in Fig.~\ref{fig:cornerplot} and the posteriors are shown as gray scale density distributions. As for the confidence intervals of the probability density function, the integral under the 2D normal distribution is given as $1 - e^{-(x/\sigma)^2/2}$, where $r$ is the distance from the mean value. The innermost contour, which refers to one standard deviation ($\sigma$), thus contains $1 - e^{-2^2/2}=39.3$\,\%. And, in analogy, the $2\sigma$ and $3\sigma$ intervals contain $86.4\%$ and $98.9\%$, respectively. With that being said, the posterior distribution that we find in Fig.~\ref{fig:cornerplot} is multimodal for most parameter pairs. Hence, the integrated probability contained inside these contours and the formal confidence intervals of our best solution do not refer to normally distributed errors. Some ground truths are formally several $\sigma$ away from the posterior peaks; however, the maximum a posteriori values are not far off from the ground truth in an absolute sense. For example, the posterior contours of $\Omega_{\rm pm}$ are clearly outside of the truth value, but the error is just $\sim1\,$\%, and this is inconsequential. The resulting light curve from the maximum likelihood solution  (Figure~\ref{fig:LC_retrieval}) is essentially a perfect match to the injected data.

\section{Discussion}

To our knowledge, {\tt Pandora} is the first photodynamical open-source exomoon transit detection algorithm. The {\tt LUNA} photodynamical modeling code by \citet{2011MNRAS.416..689K} is a proprietary code written in the {\tt FORTRAN} programming language and it has been used by the Hunt for Exomoons with Kepler survey \citep{2012ApJ...750..115K}. Our {\tt Pandora} code is open source, written in the {\tt python} programming language, and it appears to be about four to five orders of magnitude faster than {\tt LUNA}. We hope that all of this will make {\tt Pandora} accessible to a wide range of users and start a community approach in the exomoon search.

Our {\tt Pandora} code can be used with any nested sampler of choice, such as {\tt UltraNest}, {\tt MultiNest} \citep{ 2009MNRAS.398.1601F}, or {\tt Dynesty} \citep{2020MNRAS.493.3132S}.
Performance differences between nested samplers, when using the same parameters, are negligible considering a complete run. At the beginning of a sampling run, when the sampling efficiency is high ($\gtrsim 1\,$\%), a relevant fraction of computing time is spent by the sampler to identify new proposal regions. This applies to the first few percent of a search. Here, {\tt MultiNest}, written in {\tt Fortran}, is two times faster than the {\tt Python} implementations of {\tt UltraNest} and {\tt Dynesty}. Later in the run, when the sampling efficiency goes down, this fraction becomes negligible. There may exist novel sampling methods in the newer packages that could converge with fewer steps, but we have not explored them yet.

Other open-source packages such as {\tt planetplanet} \citep{2017ApJ...851...94L} and {\tt starry} \citep{2019AJ....157...64L} cover some combination of Kepler solvers and multibody occultations, but they have not been adapted for exomoon purposes. Alternative methods that analyze only the light curve that is not affected by the planetary transit \citep{2012MNRAS.419..164S,2014ApJ...787...14H,2016ApJ...820...88H,2021MNRAS.507.4120K} or TTV-TDV effects \citep{2016A&A...591A..67H} are simpler to implement, but also less sensitive \citep{2021arXiv211104444H}.

We are currently improving the fitting and sampling procedure with {\tt Pandora}, so that priors can be chosen in a more sophisticated way. To fit the quadratic limb darkening coefficients more efficiently, for example, {\tt Pandora} will offer a conversion routine to calculate $u_1=2\sqrt{q_1}q_2$ and $u_2=\sqrt{q_1}(1-2q_2)$ based on $q_1$ and $q_2$ from the unit hypercube. This procedure has been shown to reduce the prior volume \citep{2013MNRAS.435.2152K}.

The posterior distribution of our simulated exomoon search with {\tt Pandora} in Fig.~\ref{fig:cornerplot} is multimodal and it shows an interesting aliasing effect for the orbital period of the planet-moon system. This effect was first theoretically described by \citet{2009MNRAS.392..181K}. It has also been observed before in analyses of the four transits of the giant planet Kepler-1625\,b and its Neptune-sized exomoon candidate, namely in Fig.~S16 of \citet{2018SciA....4.1784T} and in Fig.~4 of \citet{2019A&A...624A..95H}. We expect this to be a general effect of Monte Carlo approaches for exoplanet-exomoon data analysis that includes a small number of transits. In other words, if only a few transits are available, there are multiple combinations of planetary masses, moon periods, and moon orbital positions that result in equally likely solutions. This degeneracy can be increasingly resolved with additional transit epochs. Some of our tests also suggest that eclipses can have an important effect on the resolution of this degeneracy. In extremely close exoplanet-exomoon systems, in which $P_{\rm pm}/2~<~d$, multiple eclipses can occur and possibly constrain the planet-moon orbital period substantially.

Beyond stellar photometry, stellar radial velocity measurements can help to constrain the total mass $M_{\rm p}+M_{\rm m}$. The simultaneous fitting of transit photometry and stellar radial velocities thus has the potential to aid further in the resolution of the parameter degeneracy.

\section{Conclusion}

{\tt Pandora} is the first open source astrophysical model software for stellar light curves with a transiting exoplanet-exomoon system. {\tt Pandora} has a short code base to increase human readability and facilitate code development, and it is optimized for computational speed. It uses well-established analytical solutions to the transit light curve with stellar limb darkening in most scenarios. In rare cases, if a planet-moon eclipse occurs at the same time when one of the two transiting bodies touches the stellar disk, {\tt Pandora} calculates the intersection of the star, the planet, and the moon numerically assuming constant stellar brightness behind the small intersecting area. {\tt Pandora} also offers the option to use the small-body approximation for the moon, which accelerates computations even further. We derived an analytical solution to the transit duration of the planet-moon barycenter in eccentric orbits around the star to avoid computationally expensive solving of Kepler's equation.

{\tt Pandora} also supports supersampling for realistic modeling of simulated observations. The output formats include light curves of the moon and the planet, as well as the combined transit model and simulated observations with arbitrary white noise amplitudes. {\tt Pandora} also comes with a video animation generator that is suitable for testing and debugging, but also for education and outreach.

Finally, we demonstrate {\tt Pandora}'s potential for exomoon searches by first simulating a physically plausible -- though not necessarily representative -- system of a Sun-like star, a Jupiter-sized transiting planet at 1\,AU from the star, and a Neptune-sized moon in a relatively tight orbit of five planetary radii in a $1.28$\,d period around the planet. We attribute a cadence and noise properties to the simulated model that corresponds to a photometrically inactive, Sun-like star as it could be observed with the 2026 PLATO mission. Our Monte Carlo search for the exomoon with {\tt UltraNest} takes about 5\,hr on a standard laptop.

Our nested sampling of simulated exoplanet-exomoon light curves reveals an aliasing effect of the orbital period in planet-moon systems. This is a fundamental aspect of transiting exoplanet-exomoon systems, which {\tt Pandora} is exquisitely suited for to explore in the future. Our next step is to search for exomoons around all of the thousands of transiting exoplanets and exoplanet candidates from the Kepler mission and to study the most interesting candidates in detail.

\begin{acknowledgements}
The authors thank the anonymous referee for a thorough report. RH acknowledges support from the German Aerospace Agency (Deutsches Zentrum f\"ur Luft- und Raumfahrt) under PLATO Data Center grant 50OO1501.
\end{acknowledgements}

\bibliographystyle{aa}
\bibliography{literature}

\appendix

\onecolumn

\section{Error estimates for constant in-transit tangential velocity approximation}
\label{sec:eccentricity_error}

The assumption that the in-transit velocity component of the planet-moon barycenter that is tangential to the celestial plane is constant in Eq.~\eqref{eq:t_dur} provides a substantial computational acceleration of {\tt Pandora} compared to a dedicated solving algorithm that approximates Kepler's equation for the eccentric anomaly. Here we estimate the error that is introduced by our approximation.

\begin{figure*}[h]
\centering
\includegraphics[width=0.495\linewidth]{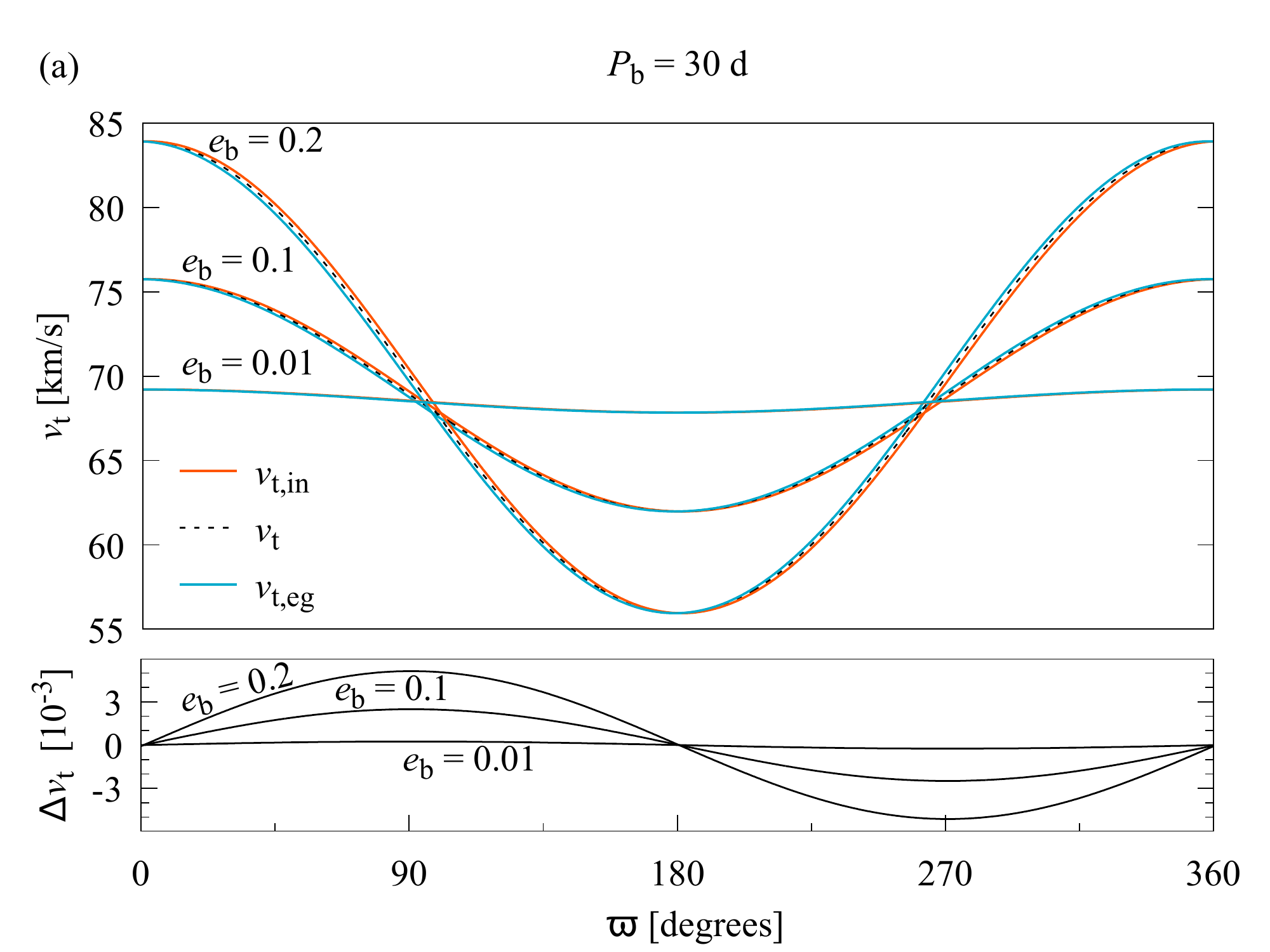}
\includegraphics[width=0.495\linewidth]{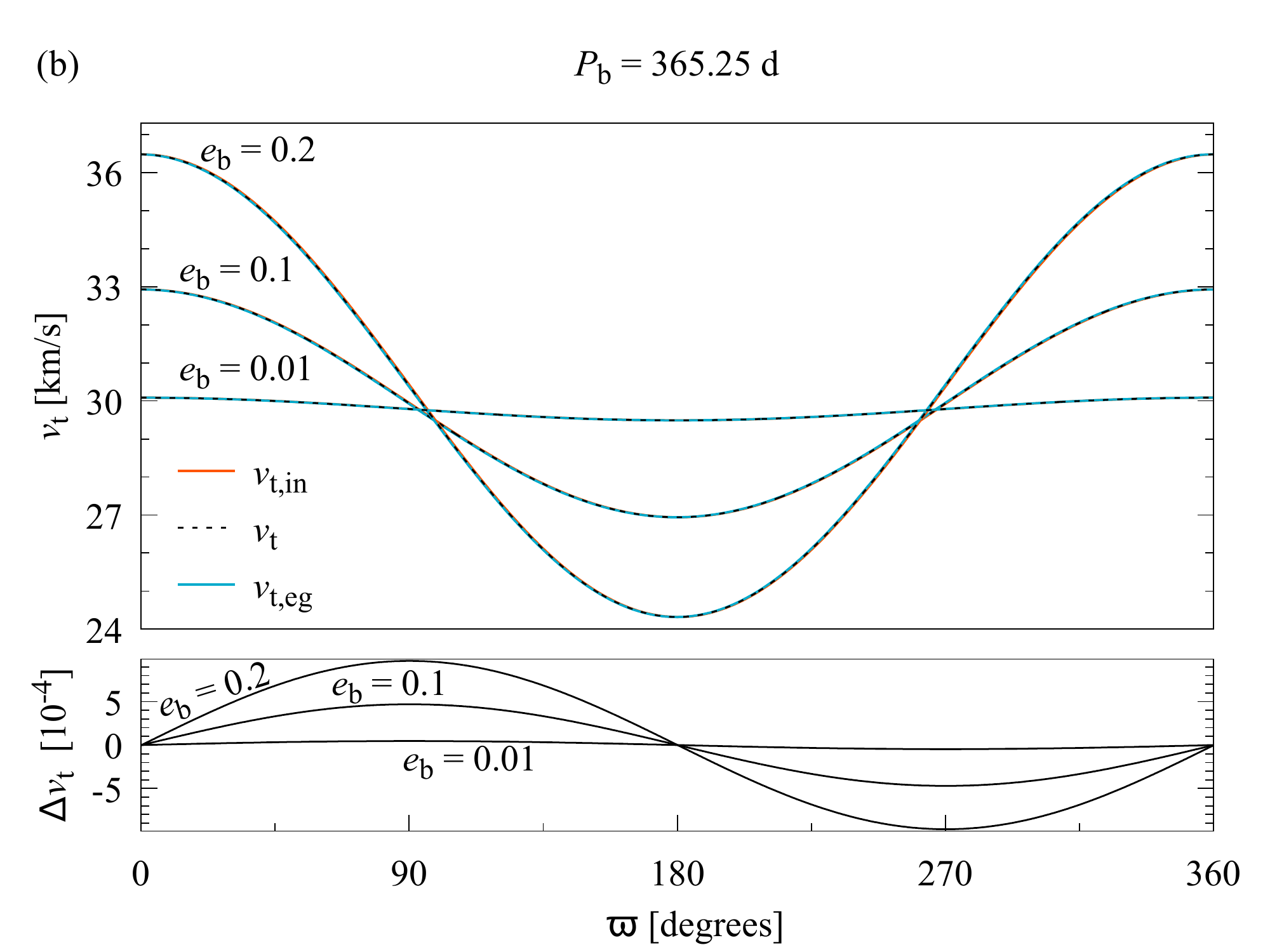}
\caption{Transit tangential velocity of the planet-moon barycenter during ingress ($v_{\rm t,in}$), midtransit ($v_{\rm t}$), and egress ($v_{\rm t,eg}$) as a function of the orientation of the periastron with respect to the line of sight ($\varpi_{\rm b}$). (a) The planet-moon barycenter has an orbital period of 30\,d around a Sun-like star. The upper panel shows $v_{\rm t,in}$ (red solid line), $v_{\rm t}$ (black dashed line), and $v_{\rm t,eg}$ (blue dashed line) for orbital eccentricities of 0.2, 0.1, and 0.01. The lower panel shows the maximum variation of the transit velocity, which we illustrate as ${\Delta}v_{\rm t}=v_{\rm t,in}/v_{\rm t,in}-1$ and in units of permille. (b) Same as (a), but with a circumstellar orbital period of 365.25\,d. The resulting variation of the in-transit tangential velocity is reduced by almost one order of magnitude, as shown in the lower panel.}
\label{fig:v_trans}
\end{figure*}

As shown in Fig.~\ref{fig:ellipse}, the true anomaly at ingress of the barycenter is $(\varpi_{\rm b} - \gamma)$, with $\gamma = \arctan(R_{\rm s}/r)$. During egress, the true anomaly is $(\varpi_{\rm b} + \gamma)$. As a consequence, the velocity component that is tangential to our line of sight during ingress and egress is

\begin{align}\label{eq:v_t_abs_in}
|\vv{v}_{\rm t,in}| = & \frac{n_{\rm b} \ a_{\rm b}}{\sqrt{1-e_{\rm b}^2}} {\Big (}1 + e_{\rm b}\cos(\varpi_{\rm b} - \gamma) {\Big )}\\ \label{eq:v_t_abs_eg}
|\vv{v}_{\rm t,eg}| = & \frac{n_{\rm b} \ a_{\rm b}}{\sqrt{1-e_{\rm b}^2}} {\Big (}1 + e_{\rm b}\cos(\varpi_{\rm b} + \gamma) {\Big ),}
\end{align}

respectively.\ In Fig.~\ref{fig:v_trans} we compare the constant velocity assumption $|v_{\rm t}|$ from Eq.~\eqref{eq:v_t_abs} with $|\vv{v}_{\rm t,in}|$ and $|\vv{v}_{\rm t,eg}|$ from Eqs.~\ref{eq:v_t_abs_in} and \ref{eq:v_t_abs_eg}, respectively. We also computed the maximum deviation between the ingress and midtransit tangential velocities, ${\Delta}v_{\rm t}~=~v_{\rm t,in}/v_{\rm t,in}-1$, as a function of $\varpi_{\rm b}$. For all curves, the stellar radius was set to $R_{\rm s}=R_\odot$ and the stellar mass to $M_{\rm s}=M_\odot$.

In Fig.~\ref{fig:v_trans}(a) we consider a planet-moon barycenter with an orbital period of 30\,d, which means that $a_{\rm b}=0.19$\,AU. The three curves assume $e_{\rm b}=0.01$, $e_{\rm b}=0.1$, and $e_{\rm b}=0.2$, respectively (see labels). The black dashed lines refer to $|v_{\rm t}|$, the orange solid line to $|\vv{v}_{\rm t,in}|$, and the light blue solid line to $|\vv{v}_{\rm t,eg}|$. The lower panel shows the resulting ${\Delta}v_{\rm t}$ values and we find that the maximum value, or error in our assumption of constant in-transit tangential velocity, reaches maximum values of about $5~\times~10^{-3}$ for $e_{\rm b}=0.2$, roughly $2.5~\times~10^{-3}$ for $e_{\rm b}=0.1$, and approximately $2.5~\times~10^{-4}$ for $e_{\rm b}=0.01$. These maximum errors occur for $\varpi_{\rm b} = 90^\circ$ and $\varpi_{\rm b} = 270^\circ$, when the change in the tangential orbital velocity is largest.

In Fig.~\ref{fig:v_trans}(b) we consider a planet-moon barycenter with an orbital period of 365.25\,d, which means that $a_{\rm b}=1$\,AU, again for $e_{\rm b}=0.01$, $e_{\rm b}=0.1$, and $e_{\rm b}=0.2$. The resulting maximum error in our assumption of constant in-transit tangential velocity is about $9.7~\times~10^{-4}$ for $e_{\rm b}=0.2$, roughly $4.7~\times~10^{-4}$ for $e_{\rm b}=0.1$, and approximately $4.6~\times~10^{-5}$ for $e_{\rm b}=0.01$.

\begin{figure}
\centering
\includegraphics[width=0.5\linewidth]{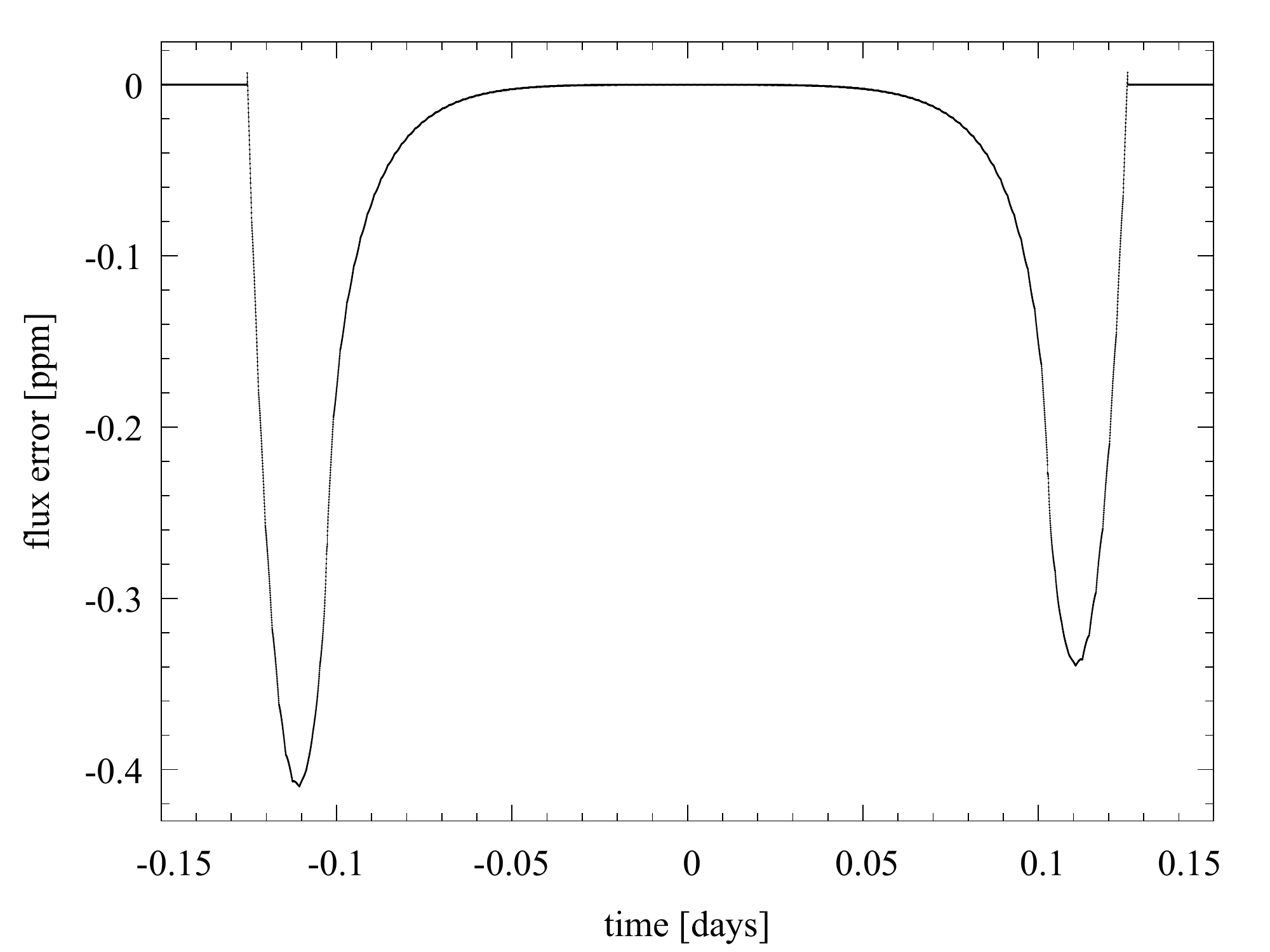}
\caption{Example of the flux error resulting from our approximation for the transit duration in an eccentric ($e=0.2$) circumstellar orbit. We assume a worst-case orientation of periastron ($\varpi_{\rm b}=90^\circ$) and a Jupiter-sized planet to exaggerate the error. The error was computed as the difference between two light curves computed with the {\tt batman} package. One light curve was initially simulated as a circular orbit and then stretched with a factor of $s = (1-e_{\rm b}^2)^{1/2} / (1 + e_{\rm b}\cos(\varpi_{\rm b}) )$ as per Eq.~\eqref{eq:t_dur}. The other light curve was computed with $e=0.2$ and using a computationally much more demanding Kepler solver for the orbital motion of the planet.}
\label{fig:ecc}
\end{figure}

We estimated the resulting error in the flux measurements for the transit light curve of a Jupiter-sized planet ($R_{\rm p}=0.1\,R_{\rm s}$) in the worst-case scenario for these above cases, in which $P_{\rm b}=30\,$d, $e_{\rm b}=0.2$, and $\varpi_{\rm b}=90^{\circ}$. We used the {\tt batman} package and its internal Kepler orbit solver to generate a highly accurate model light curve. This approach takes the variable orbital velocity during transit due to eccentricity into account. We calculated 10,000 in-transit data points, each of which is based on the accurately calculated orbital position of the planet as a function of time. For comparison, we calculated a similar light curve with {\tt batman} for $e=0$ and then stretched this light curve in time via multiplication with a factor $s = (1-e_{\rm b}^2)^{1/2} / (1 + e_{\rm b}\cos(\varpi_{\rm b}) )$ as per Eq.~\eqref{eq:t_dur} to produce the transit duration that corresponds to our assumption of constant in-transit velocity. In this particular case, we obtained $s = 0.9797959$.

Figure~\ref{fig:ecc} shows the resulting flux error as a function of time, with the origin of the time axis set to the middle of the transit. The maximum flux difference is about $4~{\times}~10^{-7}$ during ingress. We also tested the error in our approximation for different planetary radii and noticed that the error scales roughly linearly with $R_{\rm p}$. For Earth-sized transiting objects, the error is a few times $10^{-8}$.

\section{Error estimates for numerical modeling of planet-moon eclipses}
\label{sec:eclipse_error}

If a planet-moon eclipse occurs in combination with an ingress or egress of either the moon or the planet, then {\tt Pandora} switches into a numerical modeling mode, that is, to eclipse case~(2) (Sect.~\ref{sec:eclipsecase2}). The moon is modeled as a pixelated area with a diameter of $n$ pixels. The circumference of the moon then is $\pi n$ and the error of the approximated moon area is on the order of one pixel along its circumference, or $(\pi n)^{-1}$. For $n=25$, as in the default {\tt Pandora} mode, the resulting error for the moon area is ${\sim}1.3\,$\%. As an example, an Earth-sized moon transiting a Sun-like star causes a flux drop of ${\sim}84\,$ppm. The numerical error in the light curve is then ${\sim}1$\,ppm and therefore more than an order of magnitude below the astrophysical variability of a photometrically quiet star like the Sun on a time scale of minutes to hours.

\begin{figure}[h!]
\centering
\includegraphics[width=0.5\linewidth]{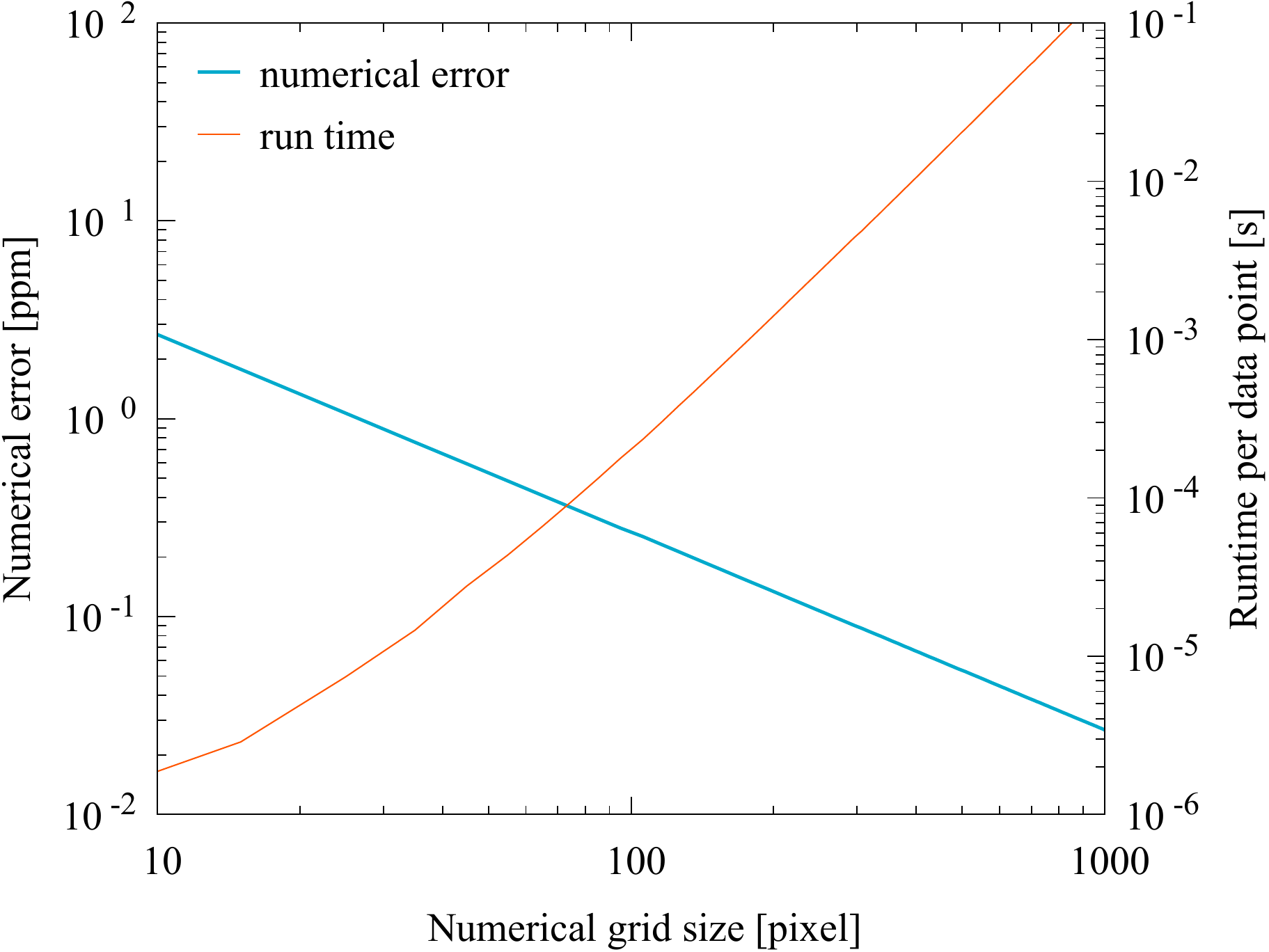}
\caption{Error estimates (blue line, left ordinate) and CPU run time per data point (orange line, right ordinate) as a function of the moon diameter in {\tt Pandora}'s numerical eclipse case (2) (see Sect.~\ref{sec:eclipsecase2}). The error refers to the absolute error in the flux of the transit light curve of an Earth-sized body in front of a Sun-like star. The error falls below 1\,ppm for grid sizes $>25$\,px, which is {\tt Pandora}'s default mode. Users of {\tt Pandora} can choose suitable grad sizes depending on accuracy requirements.}
\label{fig:grid}
\end{figure}

In addition to these analytical estimates, we undertook a detailed suite of numerical simulations with {\tt Pandora} and compared the area of a hypothetical Earth-sized moon from the pixelated area in {\tt Pandora}'s eclipse case (2) to the actual moon area. The resulting error and run time per data point are shown in Fig.~\ref{fig:grid}. The blue line illustrates that the error in the light curve is 1\,ppm for a moon diameter of 25 pixels, which is in perfect agreement with our theoretical estimates. The orange line denotes that the resulting computer run time is $7.4\,{\mu}s$ per data point.

As for the frequency and duration of planet-moon eclipses, they are rare in the Solar System and they can be expected to be rare in most exomoon scenarios. The highest frequency and longest duration occurs in an edge-on geometry, when the planet-moon orbit is in alignment with the line of sight. In this case, the maximum fraction of the planet-moon orbit spent in eclipse is roughly $2R_{\rm p}/(\pi a_{\rm pm})$. For the Earth-Moon system, this corresponds to about 1\,\% of the time; also, for the Galilean moons around Jupiter, this corresponds to about 10.4\,\% (Io), 6.6\,\% (Europa), 4.1\,\% (Ganymede), and 2.3\,\% (Callisto), respectively.

The fraction of planet-moon eclipses which happens during ingress or egress, that is, {\tt Pandora}'s eclipse case (2) that we solved numerically, is even smaller. For an Earth transit across the diameter of the Sun, ingress and egress account for 1.8\,\% of the total transit duration. For Jupiter, it is 20\,\%. Consequently, the average time for which a planet-moon occultation occurs during ingress or egress is only 0.018\,\% of the transit data points for an Earth-Moon system, or 2.1\,\% (Io), 1.3\,\% (Europa), 0.82\,\% (Ganymede), and 0.46\,\% (Callisto) for the Galilean moons, on average.

\section{Algorithmic insights for higher speed}
\label{sec:algo}

For computational speed optimization of {\tt Pandora}, it has proven crucial to break the code down into modules that can be benchmarked separately. It is often hard to estimate whether a change to the code makes it slower or faster, but speed can -- and in our opinion should -- be measured.

\subsection{Technical implementation}

All modules were implemented in pure {\tt python}, so {\tt Pandora} does not require any compilation of {\tt C} or {\tt Fortran} code. {\tt Pandora} also uses just-in-time (JIT) compilation with {\tt numba}, which reduces {\tt Pandora}'s run time by a factor of about 1100. All numbers with respect to the performance of {\tt Pandora} given in this paper are {\tt numba}-based.

To further assess the performance of our implementation, we rewrote our algorithm for the Kepler ellipse in the circular case in the {\tt C} language. The pure-{\tt C} version, compiled with GCC-11.1, is equally fast within our measurement errors of a few percent. Similarly, we benchmarked our occultation code against {\tt batman} ({\tt C} language), {\tt ALFM} \citep[{\tt C++} and {\tt Julia}]{2020AJ....159..123A}, and {\tt PyTransit} (also in {\tt python}-{\tt numba}). Depending on the parameter space, we find {\tt Pandora} to be 5--20\,\% faster. In summary, the attractiveness of {\tt Pandora} is in our combination of JIT compilation for optimal speed and {\tt python} as a programming language to ensure readability by a wide range of users.

{\tt Pandora} first determines the spatial coordinates for all points, and afterwards all occultation fluxes, etc. This allows each module to iterate over all values sequentially, making full use of a CPU core's pipeline and cache, resulting in a $\sim 10 \times$ performance gain. We have also optimized {\tt Pandora} to make use of instruction-level parallelism in modern superscalar processors. In each clock cycle, multiple instructions can be executed, increasing the throughput by a factor of a few. A key requirement to achieve this is the reduction of conditional ({\tt if}) statements.

\begin{figure*}
\centering
\includegraphics[width=0.47\linewidth]{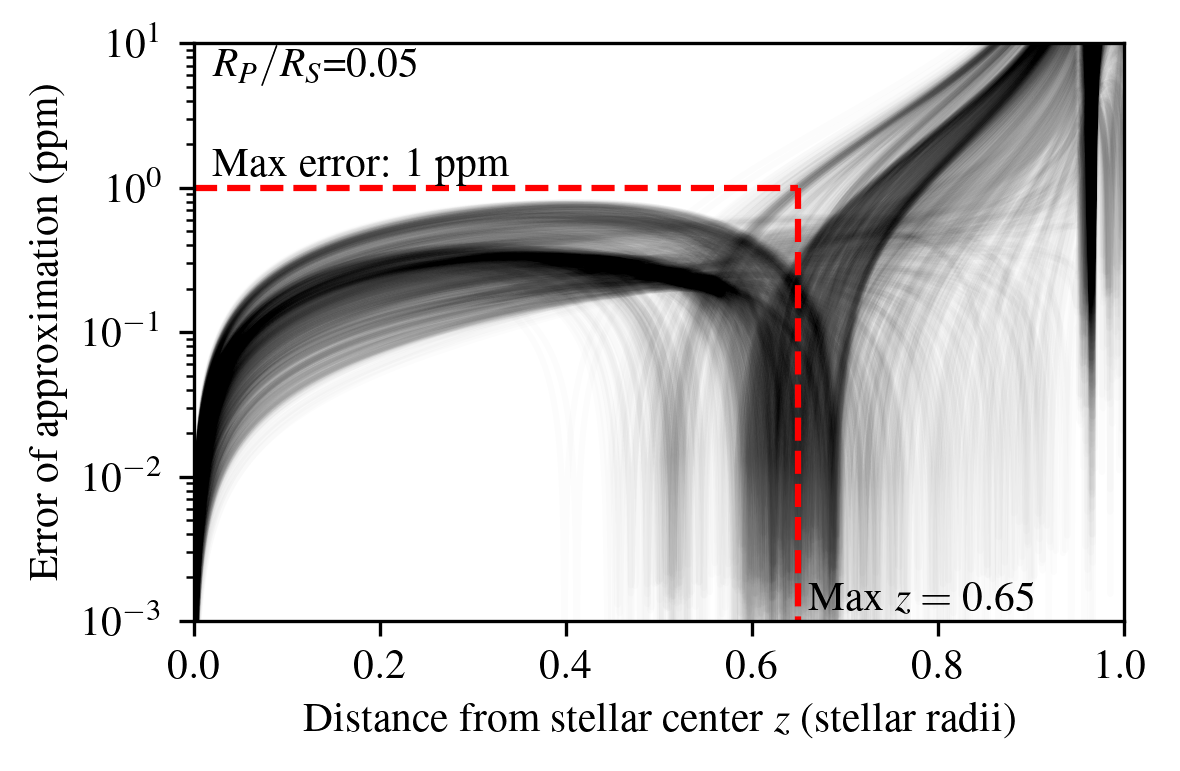}
\includegraphics[width=0.47\linewidth]{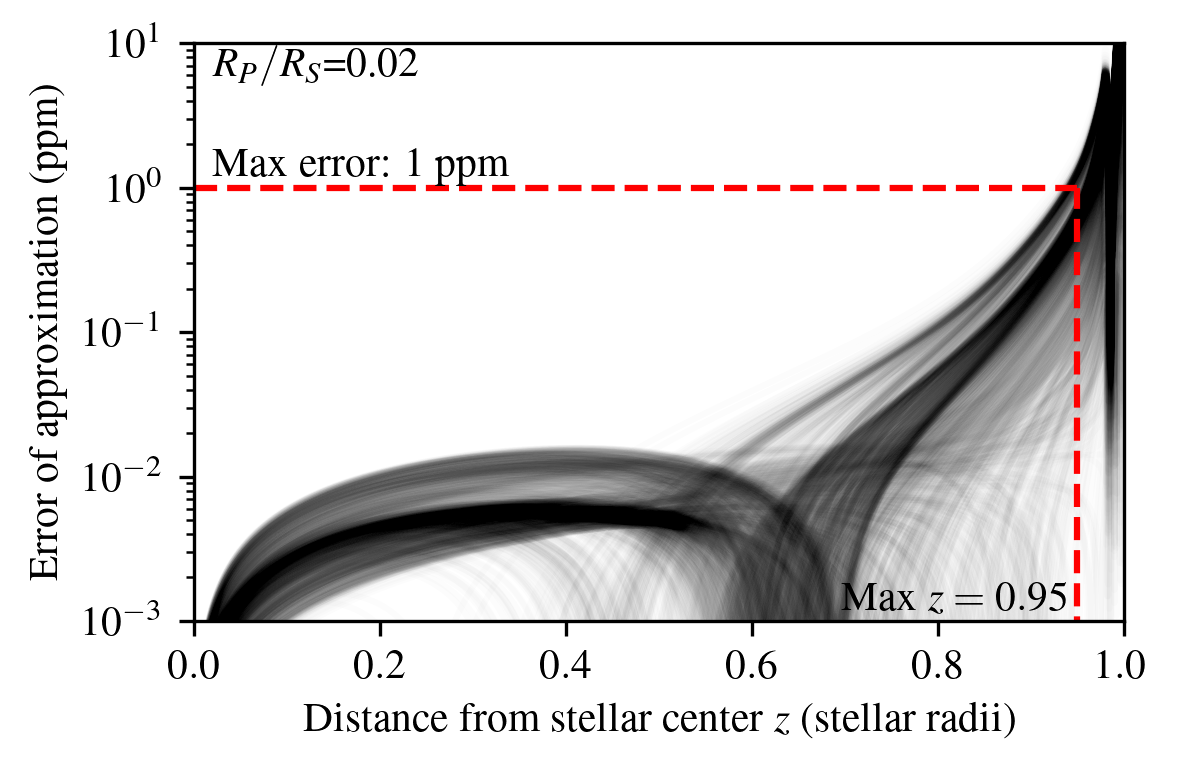}
\caption{Flux errors resulting from the small-body approximation with linear interpolation. {\it Left}: For $R_{\rm p}/R_{\rm s} = 0.05$, the maximum distance between the center of the stellar disk and the occulting body that provides errors $<1$\,ppm is $z=0.65$. {\it Right}: For $R_{\rm p}/R_{\rm s}=0.02$, flux errors are $<1\,$ppm up to $z=0.95$. Black curves are realizations for $1400$ optical limb darkening coefficients of all stellar types \citep{2017A&A...600A..30C}.}
\label{fig:small_body_approximation}
\end{figure*}

\begin{figure}
\centering
\includegraphics[width=.5\linewidth]{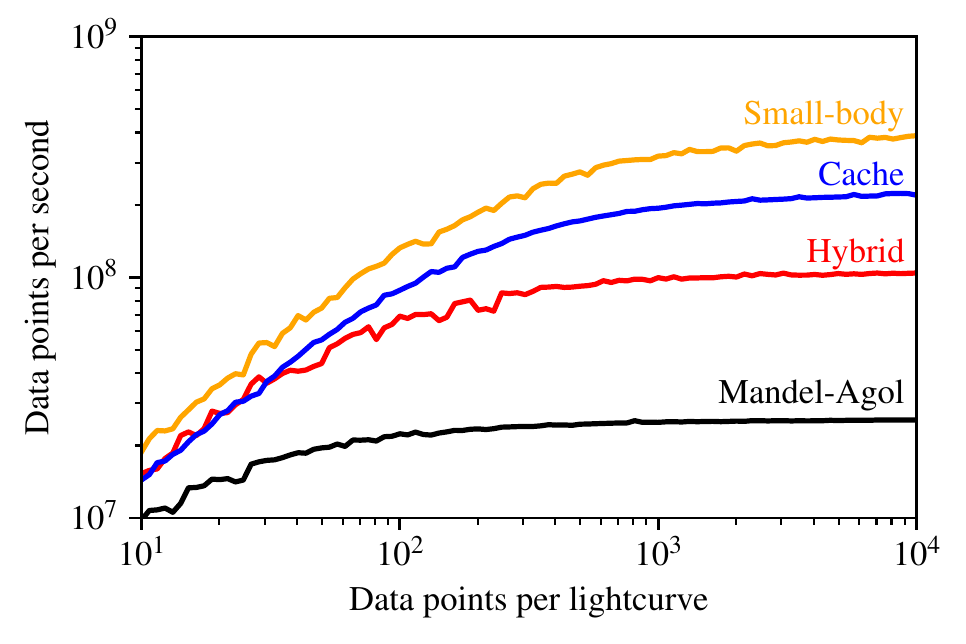}
\caption{Performance comparison of the four methods provided by {\tt Pandora} to calculate occultation fluxes. The Mandel-Agol model (black) is the most precise, but also slowest, reference version. The hybrid model (red) uses interpolation between the small-body and the Mandel-Agol methods, and it is applicable for $R_P/R_S<0.05$ to keep errors below 1\,ppm. The cached algorithm (blue) is still faster, but only usable for sampling with fixed limb darkening. Finally, the small-body approximation is the fastest method, but only applicable (errors below 1\,ppm) for $R_P/R_S<0.01$. All methods show better performance for longer light curves due to constant factors. Numbers are given for a single core on an Intel Core-i7 1185G.}
\label{fig:occult_speed}
\end{figure}

\subsection{Conditional small-body approximation with linear interpolation}
\label{appendix:small_body_approxi}
The small-body approximation assumes constant limb darkening under the occulted area when calculating fluxes. It is $\sim16\times$ faster to calculate than the exact Mandel-Agol equations. Errors from this approximation are most pronounced during ingress and egress, and they become very small ($<$\,ppm) for bodies $R_{\rm p}/R_{\rm s} < 0.01$. Larger bodies suffer from relevant errors during most of the transit, and the size of this error is dependent on the limb darkening coefficients. To reduce this error, we performed a linear interpolation between the exact and the approximated functions. This requires only the calculation of two exact values, which takes a negligible amount of time, and allows one to determine most of the occultation light curves with the small-body approximation. For $R_{\rm p}/R_{\rm s} = 0.05$, the approximation can be used for 65\,\% of the data points in a light curve, growing to 95\,\% for  $R_{\rm p}/R_{\rm s} = 0.02$, keeping small $<1\,$ppm errors (Figure~\ref{fig:small_body_approximation}). For bodies $R_{\rm p}/R_{\rm s} > 0.05$, the method is not useful because errors become significant.

The interpolation method requires one to calculate the exact flux $F_x(z, k, u_1, u_2)$ for only two radial distances $z=0$ and $z=0.65$, the latter being determined empirically ($k$ is the occulter radius in units of $R_{\rm p}/R_{\rm s}$, and $u_1$ and $u_2$ are the quadratic limb darkening coefficients). We then calculated all other values $z<z_{\rm cutoff}$ with the small-body approximation and subtracted the difference between these values and the linear interpolation through $F_1$ and $F_2$ to reduce the errors. We chose the cutoff so that errors remained $<1\,$ppm, allowing for
$z<0.65$ ($k=0.05$),
$z<0.70$ ($k=0.04)$,
$z<0.80$ ($k=0.03$),
$z<0.95$ ($k=0.02$), and
$z<0.98$ ($k=0.01$).
We hard-coded these values in Pandora. The computational performance of these methods, as a function of the number of datapoints, is shown in Fig.~\ref{fig:occult_speed}.

\subsection{Substitution of trigonometric functions}
\label{sec:trigo}

Through extensive profiling, it became clear that the major computational cost (90\,\%) for the Kepler ellipse calculation was originally in the calculation of trigonometric functions. Standard implementations such as the one in {\tt PyAstronomy} calculate a range of $\sin$, $\cos$, $\tan$, and $\arctan$ terms for each resulting data point, where the 3D ($x,y,z$) positions are determined via

\begin{align}
k &= \pi ({\rm time} - \tau) / {\rm per} \\
Q &= 2 \arctan(\tan(k)) \\
V &= \sin(Q) \cos(i) \\
x &= (\cos(\Omega) \cos(Q) - \sin(\Omega) V)\, a \\
y &= (\sin(\Omega) \cos(Q) + \cos(\Omega) V)\, a \\
z &= \sin(Q) \sin(i)\,a.
\end{align}

First, we can substitute $Q=\arctan(\tan(k)) \equiv k$. Then, profiling shows that almost all time is spent on the expressions $\sin(2 k)$ and $\cos(2 k)$ (where $k=\pi ({\rm time} - \tau) / {\rm per})$, which is trivially fast to calculate). As trigonometric functions are very expensive, we can substitute the sine and cosine calculations for one tan calculation through the following identity:

\begin{align}
\cos(2k) = \frac{1 - \tan^2(k)}{1 + \tan^2(k)}
\end{align}

\noindent
and

\begin{align}
\sin(2k) = \frac{2 \tan(k)}{1 + \tan^2(k).}
\end{align}

On standard CPU architectures (Intel Core i7 and Apple M1), the substitution is $\sim10$\,\% faster.
Additional trigonometry is still formally required to determine the Kepler ellipse, but only once, and not for each of the $\sim{1000}$ data points on the ellipse orbit. That is to say, each data point can be calculated with only one trigonometric calculation of $\tan(k)$. Finally, we can remove the calculation of the depth component $z$ because the orientation of the planet-moon system is irrelevant as both are black disks.

\subsection{Numerical accuracy requirements}

In computer hardware, floating-point numbers are represented as base-2 (or ``binary'') fractions. With a finite number of bits available for a given number, any number is an approximation. High-level programming languages such as {\tt python} use double-precision floating point data types with 64-bit accuracy, resulting in almost 16 significant decimal digits of precision. The standard math library follows the IEEE-754 arithmetic, which aims to provide accuracy to the last significant digit, that is, an error $<0.5$ in the unit in the last place (ULP). As errors can add up over the course of multiple calculations, it is useful to test the requirements for a given problem. With that being said, it appears to us that calculating every value in {\tt Pandora} to 16 digits would be excessive..

In the case of the Kepler ellipse, we calculated a tangent for each point on the orbit and then multiplied this value with various other parameters. We required an accuracy well below the noise present in the data, which is typically $10^{-5}$. The signal of Earth's Moon transiting a Sun-like star would be about 6\,ppm ($6\times10^{-6}$), which implies a precision requirement of at least seven decimal digits to achieve a model that is substantially more accurate than the signal. This requirement is about nine orders of magnitude weaker than the 16 digits typically available in our {\tt python} implementation of {\tt Pandora}.

There are two ways to reduce computational effort at the cost of accuracy: Single precision floats and/or a different trigonometric algorithm. The {\tt numba} package is built upon the {\tt LLVM} compiler stack, which offers a set of {\tt fastmath} approximations, resulting in up to four ULPs of errors. The speed gain with this option is 20--50\,\%. Single precision floats provide 7.225 decimal digits of precision on average, which is insufficient given our requirement of seven digits and multiple calculations on intermediate values. The speed difference would be a factor of two on most architectures. Working with double precision (16 digits), we can accept many ULPs of fast approximation errors. Consequently, {\tt Pandora} uses 64-bit arithmetic, but low accuracy (four ULPs) approximations.

\subsection{Potential for future algorithmic optimization}

Various alternative algorithms exist which could further improve {\tt Pandora}'s performance. We have deferred the implementation of these for future releases, given sufficient interest by the community.

As an example, our Kepler solver for the eccentric case uses the method by \citet{1995CeMDA..63..101M}. Faster algorithms exist, for example one is based on a CORDIC-like solver \citep{2018A&A...619A.128Z,2021MNRAS.500..109Z}. As eccentric moons are rarely expected, we have deferred to these optimizations. 

The occultation algorithm {\tt ALFM} \citep{2020AJ....159..123A} uses a different method to calculate the circle-circle intersect than the one used  by {\tt Pandora}. The standard method of computing the overlap of two circles requires the calculation of a square-root and of two $\arccos$ terms. The new ``kite expression'' replaces the $\arccos$ with inverse $\arctan$, keeping the same constant factors, while achieving higher numerical accuracy. Our benchmarks in {\tt numba} and {\tt C}-language yield identical performance within a measurement uncertainty of 1\,\%. Due to the lower complexity of the standard approach, we decided to keep the classical circle-circle intersect formula in {\tt Pandora} for now. The improved accuracy of the kite expression is irrelevant for our purpose, as the errors from the standard method are $<10^{-8}$ in all cases \citep[Fig.~2 in][]{2020AJ....159..123A}.

The occultation part of {\tt Pandora} also contains code to calculate the complete elliptical integral of the third kind. A well-known standard method for this calculation, which is also used in {\tt EXOFAST} \citep{2013PASP..125...83E}, is the iterative approximation by \citet{Bulirsch1965,Bulirsch1965b}. As an alternative, \citet{Fukushima2013} describes a method based on half and double argument transformations, which is benchmarked by the author to be $20-50$\,\% faster than Bulirsch's method. We are grateful to Toshio Fukushima for sharing his {\tt Fortran} code ($\sim 700$ lines of code, including precalculated constants) to perform a comparison. In our use case, however, our {\tt numba} version of Bulirsch's method is slightly faster.

\subsection{GPU-based calculations}

We have implemented a {\tt CUDA} version of {\tt Pandora} to run on Nvidia graphics cards (GPUs). A performance test on a Nvidia GTX 1080 card, with $3584$ Cuda-cores, yields a sustained throughput of $225{,}000$ model calculations per second, including the corresponding log-likelihood evaluations. This is a speed-up by a factor of 20 compared to a single core of an Intel Core i7-7700k processor. Unfortunately, it appears that current versions of nested samplers such as {\tt UltraNest} cannot process this load sufficiently fast and that they are limited to $\sim50{,}000$ point proposals per second due to internal overhead. We plan to address these issues in a subsequent version of {\tt Pandora} because a 20-fold performance gain would reduce typical convergence times from five hours to fifteen minutes.

\section{Usage examples}

\subsubsection{Defining model parameters}

\begin{python}
import pandoramoon as pandora
params = pandora.model_params()
params.per_bary = 365.25  # [days]
params.a_bary = 215       # [R_Star]
params.r_planet = 0.1     # [R_Star]
params.b_bary = 0.3       # [0..2]
(...)
\end{python}
A full list of model parameters is available in the online documentation.

\subsubsection{Creating a light curve}
\begin{python}
time = pandora.time(params).grid()
model = pandora.moon_model(params)
flux_total, flux_planet, flux_moon = 
    model.light_curve(time).
\end{python}

\subsubsection{Obtaining coordinates}
The return values $px$, $py$ determine the $x$ and $y$ components of the planet, while $mx$ and $my$ are for the moon; they were sampled at each time stamp.

\begin{python}
px, py, mx, my = model.coordinates(time)
\end{python}

\subsubsection{Generating a transit animation video}
\begin{python}
video = model.video(
    time,
    limb_darkening=True, 
    teff=3000, 
    planet_color="black",
    moon_color="black",
    ld_circles=100
)
video.save(
   filename="video.mp4",
   fps=25, 
   dpi=300
).
\end{python}

\end{document}